\documentstyle[aas2pp4,epsfig,flushrt,twoside]{article}
\renewcommand{\vec}[1]{\mbox{\protect\boldmath $\displaystyle #1$}} 
\newcommand{\grad}{\vec{\nabla}} 
\newcommand{\vdot}{\vec{\cdot}} 
\newcommand{\vcross}{\vec{\times}} 
\newcommand{\divr}{\grad\vdot\,} 
\newcommand{\curl}{\grad\vcross\,} 
\newcommand{\ddt}[1]{\frac{\partial #1}{\partial t}} 
\newcommand{\DDt}[1]{\frac{d #1}{dt}} 
\newcommand{\ddy}[1]{\frac{\partial #1}{\partial y}} 
\newcommand{\DDy}[1]{\frac{d #1}{dy}} 
\newcommand{\ddz}[1]{\frac{\partial #1}{\partial z}} 
\newcommand{\kB}{k_{\rm B}{}}	
\newcommand{\pF}{p_{\rm F}{}}	
\newcommand{\K}{\,{\rm K}}	
\newcommand{\gram}{\,{\rm g}}	
\newcommand{\second}{\,{\rm s}}	
\newcommand{\minute}{{\rm\,min}} 
\newcommand{\hour}{{\rm\,hr}}	
\newcommand{\yr}{\,{\rm yr}}	
\newcommand{\cm}{\,{\rm cm}}	
\newcommand{\km}{\,{\rm km}}	
\newcommand{\GramPerCc}{\gram\cm^{-3}} 
\newcommand{\GramPerSc}{\gram\cm^{-2}} 
\newcommand{\erg}{\,{\rm erg}} 
\newcommand{\MeV}{\,{\rm MeV}}	
\newcommand{\keV}{\,{\rm keV}}	
\newcommand{\gauss}{\,{\rm G}}	
\newcommand{\msun}{M_{\odot}}	
\newcommand{\mdot}{\dot{m}}	
\newcommand{\medd}{\mdot_{\rm Edd}} 
\newcommand{\gp}{{\cal V}}	
\newcommand{\tfl}{t_{\rm fl}}	
\newcommand{\tth}{t_{\rm th}}	
\newcommand{\ecool}{\epsilon_{\rm cool}} 
\newcommand{\tnu}{t_\nu}	
\newcommand{\enu}{\epsilon_\nu}	
\newcommand{\tgh}{t_{\rm gh}}	
\newcommand{\enuc}{\epsilon_{\rm nuc}} 
\newcommand{\delab}{\nabla_{\rm\!ad}} 
\newcommand{\mcrit}{\mdot_{\rm crit}} 
\newcommand{\localrate}{\GramPerSc\second^{-1}} 
\newcommand{\ee}[1]{\times 10^{#1}} 
\newcommand{\nuc}[1]{$\rm #1$}	
\newcommand{\C}{\nuc{^{12}C}}	
\newcommand{\Ox}{\nuc{^{16}O}}	
\newcommand{\Ne}{\nuc{^{20}Ne}} 
\newcommand{\Mg}{\nuc{^{24}Mg}} 
\newcommand{\Si}{\nuc{^{28}Si}} 
\newcommand{\alphagamma}{\nuc{\,(\alpha,\gamma)\,}} 
\begin{document}
\pagestyle{myheadings}
\markboth{Brown \& Bildsten}{Ocean and Crust of a Rapidly 
	Accreting Neutron Star}
\title{The Ocean and Crust of a Rapidly Accreting Neutron Star:
Implications for Magnetic Field Evolution and Thermonuclear Flashes}
\author{Edward F. Brown and Lars Bildsten}
\affil{ Department of Physics and Department of Astronomy\\
	601 Campbell Hall, Mail Code 3411, University of California,
	Berkeley, CA 94720--3411\\
	e-mail: ebrown@astron.berkeley.edu, bildsten@fire.berkeley.edu}
\thispagestyle{empty}

\vspace{10pt}
\centerline{\large To appear in {\sc the Astrophysical Journal}, 1 April 1998}
\vspace*{-30pt}

\begin{abstract}

We investigate the atmospheres, oceans, and crusts of neutron stars
accreting at rates sufficiently high (typically in excess of the local
Eddington limit) to stabilize the burning of accreted hydrogen and
helium.  For hydrogen-rich accretion at global rates in excess of
$10^{-8}\msun\yr^{-1}$ (typical of a few neutron stars), we discuss the
thermal state of the deep ocean and crust and their coupling to the
neutron star core, which is heated by conduction (from the crust) and
cooled by neutrino emission.  We estimate the Ohmic diffusion time in
the hot, deep crust and find that it is noticeably shortened (to less
than $10^8\yr$) from the values characteristic of the colder crusts in
slowly accreting neutron stars.  As suggested by Konar \& Bhattacharya,
at high accretion rates the flow timescale competes with the Ohmic
diffusion time in determining the evolution of the crust magnetic field.
At a global accretion rate of $\dot{M}\approx 3\times 10^{-9}
\msun\yr^{-1}$, the Ohmic diffusion time across a scale height equals
the flow time over a large range of densities in the outer crust.  In
the inner crust (below neutron drip), the diffusion time is always
longer than the flow time, for sub-Eddington accretion rates.  We
speculate on the implications of these calculations for magnetic field
evolution in the bright accreting X-ray sources.

We also explore the consequences of rapid compression at local accretion
rates exceeding ten times the Eddington rate.  This rapid accretion
heats the atmosphere/ocean to temperatures of order $10^9\K$ at
relatively low densities; for stars accreting pure helium, this causes
unstable ignition of the ashes (mostly carbon) resulting from stable
helium burning.  This unstable burning can recur on timescales as
short as hours to days, and might be the cause of some flares on helium
accreting pulsars, in particular 4U~1626--67.  Such rapid local
accretion rates are common on accreting X-ray pulsars, where the
magnetic field focuses the accretion flow onto a small fraction of the
stellar area.  We estimate how large such a confined ``mountain'' could
be, and show that the currents needed to confine the mountain are large
enough to modify, by order unity, the magnetic field strength at the
polar cap.  If the mountain's structure varies in time, the changing
surface field could cause temporal changes in the pulse profiles and
cyclotron line energies of accreting X-ray pulsars.

\end{abstract}


\section{Introduction}\label{s:introduction}

\begin{figure}[hbt]
\centering{\epsfig{file=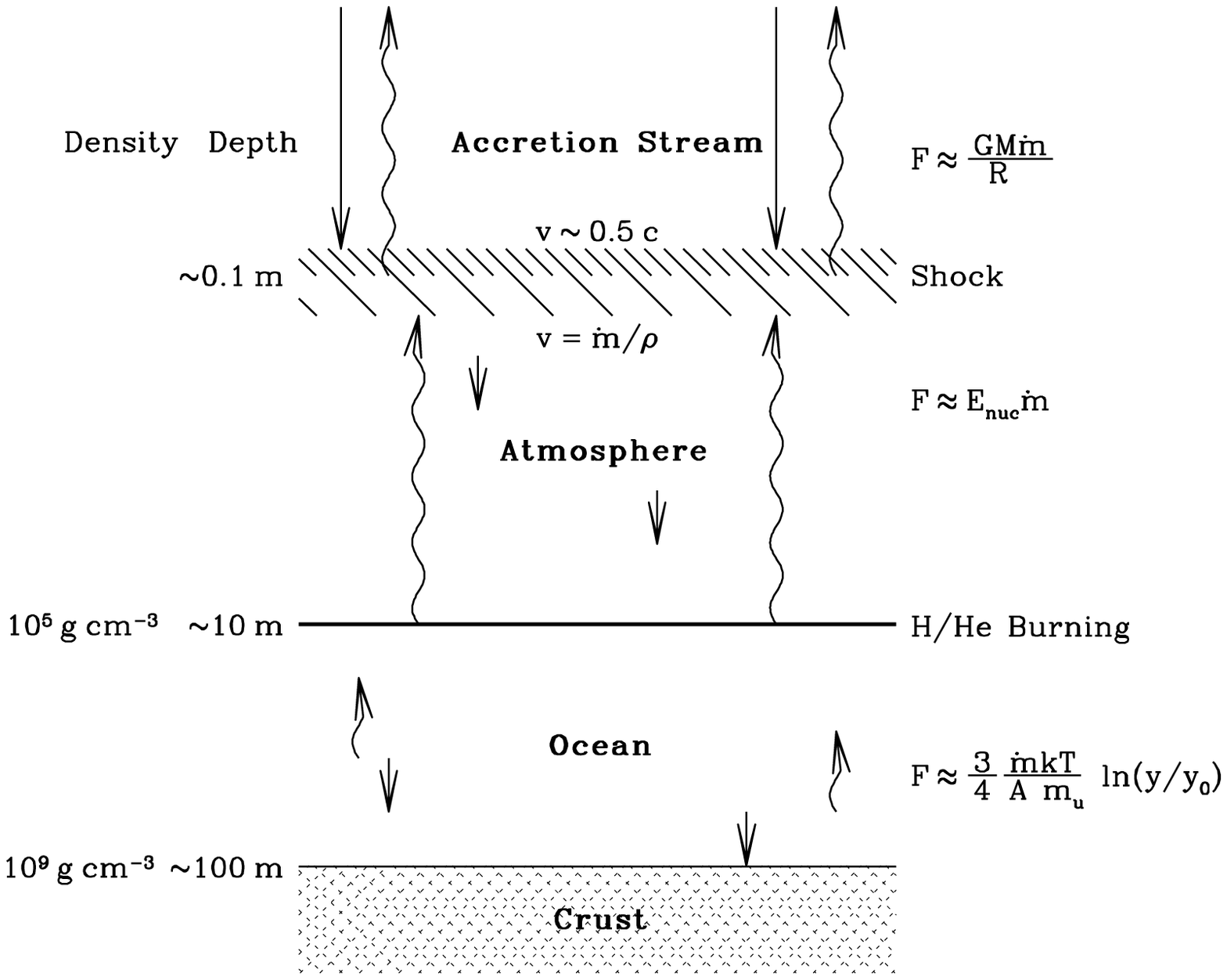,width=\hsize}}
\large\renewcommand{\baselinestretch}{0.7}
\footnotesize
\caption{\protect{\footnotesize Schematic of the outer layers of an
accreting neutron star.  Infalling matter ({\em straight arrows\/}) is
decelerated from a velocity $v\sim 0.5 c$ at a shock and radiates a flux
$F\approx GM\mdot/R$.  Beneath the shock, the accreted matter forms an
atmosphere/ocean in hydrostatic balance where the downward flow velocity
is $v=\mdot/\rho$ (\S~\ref{s:equations}).  We indicate both the density
and the physical depth (roughly the pressure scale height) of the
underlying layers on the left-hand side of the diagram.  The flux ({\em
wavy arrows\/}) in this region is set by the steady-state hydrogen and
helium burning \{here $E_{\rm nuc} = X[(Q_H/4) + (Q_{3\alpha}/12)] +
Y(Q_{3\alpha}/12)$, see eq.~[\protect\ref{e:fluxhyd}]\}.  Below the H/He
burning region, the matter forms an electron-degenerate ocean (\S\
\ref{s:DeepOcean}), where energy transport is mediated by electron
conduction.  The ocean is nearly isothermal, and the flux is set by
further compression of the matter (see eq.~[\ref{e:analyticF}]).  At a
density of roughly $10^9\GramPerCc$, the ions crystallize into a
lattice.  The thermal state of the crust and core is described in
\S~\ref{s:core}.
\label{f:schematic}}}
\end{figure}

The thermal and compositional structure of the outer layers of an
accreting neutron star is of interest to studies of accretion-induced
magnetic field decay (e.g., \cite{urp95}), thermonuclear processes, and
the thermal and compositional structure of the inner crust and core.  In
this paper, we consider objects accreting at rates high enough for
stable burning of hydrogen and helium (typically requiring
super-Eddington rates; for a recent review see Bildsten 1998) and
examine the thermal and compositional structure of the underlying ocean
and crust.  By ``ocean and crust'' we mean the region far below where
the accreted matter is decelerated from its free-fall velocity and
directly underneath the hydrogen/helium burning layer (Fig.\ 
\ref{f:schematic} provides a sectional cut through this part of the
neutron star).  We are therefore considering fluid layers that are
gradually being compressed by the continuous addition of fluid at the
top of the atmosphere.  Previous works (\cite{aya82}; \cite{fuj84};
\cite{mir90}) have examined this problem for stars accreting at globally
sub-Eddington rates in a spherically symmetrical fashion.  In this
paper, we do not impose such a restriction and therefore parameterize
accretion by the local rate per unit area, $\mdot$.  In general, neutron
stars accrete at super-Eddington rates by means of an optically thick
column; on accretion-powered X-ray pulsars, the local accretion rate in
this column can easily exceed the Eddington rate by orders of magnitude.

Our motivation is twofold.  For hydrogen-rich accretion at global rates
of order $10^{-8}\msun\yr^{-1}$ (the Eddington limit), we find the
minimum temperature in the deep crust and the resulting maximum possible
Ohmic diffusion time there.  This is applicable to the brightest
accreting objects in our galaxy (e.g. Sco X-1 and the GX objects, which
are weakly magnetic, $B\ll 10^{12}\gauss$) and the Magellanic Clouds.
For accreting pulsars with high magnetic fields ($\gtrsim
10^{12}\gauss$), we mostly consider the conditions underneath the
accreting polar cap, where magnetic confinement of the accreted matter
ensures a locally high accretion rate that rapidly compresses the ocean
and heats it to temperatures of order $10^9\K$ at low densities.  If
this accreted matter contains an appreciable amount of carbon (or
oxygen), a violent thermonuclear instability occurs and produces a flare
from the polar cap.  At super-Eddington local accretion rates, the
recurrence time for this instability is short enough (hours to days) to
allow observations of the flares.

We start in \S~\ref{s:equations} by deriving the Eulerian fluid
equations for a plane-parallel geometry.  Presuming steady-state
hydrogen/helium burning (valid for super-Eddington accretion), we
briefly investigate, in \S~\ref{s:He_burn}, the thermal structure of the
star's upper atmosphere.  Because the case of pure helium accretion is
simpler, and because it is relevant to a number of binaries, we first
study the steady-state burning of helium and show that the ashes of this
burning are mostly carbon.  When the neutron star is accreting a
hydrogen-rich mixture at approximately the Eddington rate, the stable
hydrogen burning is never fast enough to consume all of the hydrogen
before the helium ignites.  Rapid proton captures onto the rapidly
proliferating seed nuclei then produce a mix of heavy (trans-iron)
elements.  Although the final mix is quite complicated (and has yet to
be calculated), the thermal profile is readily found.

We next move down to the degenerate electron ocean, where the density is
greater than $10^6\GramPerCc$.  The thermal flux flowing through the
ocean is mostly generated by settling and compression within it.  We
show in \S~\ref{s:DeepOcean} how $\mdot$ determines the thermal profile
of this deep ocean.  Non-equilibrium nuclear reactions (mostly neutron
emissions and pycnonuclear reactions) are an important heat source in
the deep crust (\cite{hae90}).  Using a crude formulation of the crust
heat sources, we show that most of this energy is conducted into the
neutron star core, which finds a steady-state core temperature hot
enough to balance this heating with neutrino losses.

In \S~\ref{s:OhmicDecay}, we use the thermal profiles in the crust to
estimate the Ohmic diffusion time there.  The effect of
accretion-induced heating on the Ohmic decay of crust magnetic fields
has previously been considered (\cite{rom90}; \cite{gep94};
\cite{pet95}; \cite{urp95}; \cite{urp96}; \cite{kon97}) for stars
accreting at sub-Eddington rates.  In the deep crust, the local
accretion rate is roughly independent of latitude; there are, however, a
few systems that accrete {\em globally\/} at or near the Eddington rate.
As a first cut, we compute the Ohmic diffusion time across a scale
height throughout the outer crust ($\rho\lesssim 10^{13}\GramPerCc$) and
find that it is greater than the time for matter to flow across a scale
height for near-Eddington accretion rates.  This implies that crust
magnetic fields are advected to great depths before they are able to
decay.

Although variations with latitude in the local accretion rate may be
negligible deep within the crust, they cannot be ignored in the upper
ocean and atmosphere.  In the case of accretion-powered pulsars, for
which the infalling accretion stream is funneled onto a small area about
the polar cap, the local accretion rate can be very super-Eddington even
if the global accretion rate is sub-Eddington.  The lack of type I X-ray
bursts implies a locally super-Eddington accretion rate at the cap in
accreting pulsars (Joss \& Li 1980; Ayasli \& Joss 1982; Bildsten \&
Brown 1997).  In \S~\ref{s:MagneticConfinement}, we show that the matter
is tied to the field lines to very large depths and that spreading away
from the polar cap does not occur until the accreted matter reaches
pressures greatly exceeding $B^2/8\pi$ (a detailed discussion is in
Appendix \ref{s:GradShafranov}).  The resulting high local accretion
rate causes unstable carbon ignition at relatively low densities
(\S~\ref{s:ignition}).  The low ignition column depth and rapid
accretion rate combine to make the resulting large ``flares'' from the
polar caps recur on an hourly to daily timescale.  This timescale
differs strongly with the calculation of Woosley \& Taam (1976), who
found a roughly ten yr recurrence time for near-Eddington accretion
rates.  We discuss the observable signatures of such burning, and
briefly compare our theoretical models to some accreting X-ray pulsars,
including LMC~X-4 and 4U~1626--67.  We conclude in
\S~\ref{s:conclusions} with a summary of our main results.

\section{The Equations in  Planar Geometry}\label{s:equations}

Because the accretion flow onto the neutron star is in general
asymmetrical, we define a local accretion rate, $\mdot$, as the mass
accreted per unit time per unit area.  The physics of the reactions,
compression, and settling all depend on this quantity.  The local
Eddington accretion rate is
\begin{equation}
   \medd = \frac{\mu_e m_u c}{\sigma_T R} 
   	= 7.5\ee{4}\localrate\mu_e\left(\frac{R}{10\km}\right)^{-1},
\end{equation}
where $\sigma_T$ is the Thomson scattering cross-section, $c$ is the
speed of light, $R$ is the radius of the star, $m_u$ is the atomic mass
unit, and the number density of electrons is $\rho/(\mu_e m_u)$.  Future
references to $\medd$ are for $\mu_e=1$.  Because the ram pressure in
the internal sub-sonic flow is negligible, the momentum equation is just
that of hydrostatic balance,
\begin{equation}\label{e:HB}
   \frac{dP}{dz} = -\rho g, 
\end{equation}
allowing us to parameterize the spatial coordinate by column depth,
$y(z) \equiv \int_z^\infty \rho\,dz'$, so that $P(z) = gy(z)$.  The heat
flux flowing through the deep atmosphere (far beneath the photosphere,
see Fig.\ \ref{f:schematic}) is much less than the accretion flux
($\approx G\dot{M}m_u/R$), so radiation pressure is unimportant.  Since
accretion does not appreciably change the radial coordinate of isobars
(see Appendix \ref{s:isobar}), $y$ is an Eulerian coordinate.  We take
the surface gravity $g$ to be constant, since the pressure scale height,
$H=P/\rho g$, is much less than $R$ everywhere in the atmosphere.
Throughout this paper, we assume a neutron star mass and radius of
$1.4\msun$ and $10\km$, respectively; we also use Euclidean spacetime,
with $g = GM/R^2$.  We ignore components of heat transport and pressure
gradients parallel to the surface.  Even in the case of accretion onto a
small fraction of the surface, the atmospheric scale height is much
smaller than lengthscales along the surface.  As a result, the area of
the top and bottom ``faces'' of the ocean under the accretion column is
much greater than the area along the ``edges''.  For a polar cap of
radius $10^5\cm$ (a tenth of the stellar radius), the ratio of edge area
to face area is $\sim 10^{-2}$.

The continuous accretion of fresh material modifies the equations of
continuity and entropy.  The continuity equation for a species $j$ (with
number density $n_j$) is
\begin{equation}\label{e:conta}
   \ddt{n_j} + \divr (n_j\vec{v}) = 
   	-r_{\rm dest}^{(j)} + r_{\rm prod}^{(j)}, 
\end{equation}
where $r_{\rm dest}^{(j)}$ ($r_{\rm prod}^{(j)}$) is the destruction
(production) rate of species $j$ due to thermonuclear reactions.  We
presume that all ions co-move (a good approximation at high accretion
rates where the downward compression is faster than any drift time for a
heavy particle).  Defining a mass fraction
\begin{equation}
   X_j\equiv\frac{\rho_j}{\rho} = \frac{A_j m_u n_j}{\rho},
\end{equation}
where $A_j$ is the baryon number of species $j$, and, recognizing that
the accreted matter flows through constant pressure surfaces at a
velocity $v = \mdot/\rho$, we expand the continuity equation
(eq.~[\ref{e:conta}]) in $(y,t)$ coordinates to obtain
\begin{equation}\label{e:cont}
   \ddt{X_j} + \dot{m} \ddy{X_j} = 
   \frac{A_j m_u}{\rho} (-r_{\rm dest}^{(j)} + r_{\rm prod}^{(j)}).
\end{equation}
In this derivation we neglect the contribution of binding energy to
$\rho$.  The entropy equation is
\begin{equation}\label{e:entra}
   T\DDt{s} = -\frac{1}{\rho}\divr\vec{F} + \epsilon,
\end{equation}
where $\vec{F}$ is the flux and $\epsilon$ is the sum of all sources and
sinks of entropy.  The flux obeys Fick's law,
\begin{equation} \label{e:fluxa}
   \vec{F} = -K\grad T,
\end{equation}
with $K$ denoting the thermal conductivity.  Heat transport is set by
both radiative (primarily Thomson scattering and free-free absorption)
and conductive (electron-electron and electron-ion scattering)
processes.  The total thermal conductivity is then
\begin{equation}
   K = \frac{4ac}{3\rho\kappa_{\rm rad}}T^3 + K(\mbox{\rm conduction}),
\end{equation}
where $\kappa_{\rm rad}$ is the radiative opacity.  Using the equations
for hydrostatic balance (eq.\ [\ref{e:HB}]) and flow velocity
$v=\mdot/\rho$, and the thermodynamic relation
\begin{equation}
   \left(\frac{\partial s}{\partial P}\right)_T = 
	-\left(\frac{\partial T}{\partial P}\right)_s 
	\left(\frac{\partial s}{\partial T}\right)_P,
\end{equation}
we write the entropy as a function $s=s(T(y,t),P(y))$ and transform
equations (\ref{e:entra}) and (\ref{e:fluxa}) into the more useful forms
\begin{mathletters}
\begin{eqnarray}
\label{e:entropy}
\ddy{F} + \epsilon &=& 
   c_{\!P}\left(\ddt{T}+\dot{m}\ddy{T}\right)
	-\frac{c_{\!P}T\mdot}{y}\delab \\
\label{e:flux} 
   \ddy{T} &=& \frac{F}{\rho K}.
\end{eqnarray}
\end{mathletters}
Here $c_{\!P}$ is the specific heat at constant pressure, and $\delab\equiv
(\partial\ln T/\partial\ln P)_s$ is the adiabat.  Equations
(\ref{e:cont}), (\ref{e:entropy}), and (\ref{e:flux}) with appropriate
boundary conditions describe the thermal evolution of the outer layers
of the neutron star.

\section{Hydrogen and Helium Burning in the Upper Atmosphere}
\label{s:He_burn}

We now derive the thermal and compositional state of the upper
atmosphere for these characteristically high accretion rates.  This
determines the outer starting point for the derivation of the conditions
deeper in the ocean and crust.  The accreting hydrogen and helium
typically fuse to heavier elements within an hour after arrival at the
neutron star atmosphere.  Because of $\beta$-decay limitations, hydrogen
burning via the CNO cycle is stable at $\mdot \gtrsim 10^{-2} \medd $
and never fast enough to consume the accreting hydrogen before the
helium ignites (\cite{lam78}; \cite{taa79}; Fujimoto, Hanawa, \& Miyaji
1981).  For $\mdot \gtrsim 10^{-1} \medd$, helium burning occurs in a
hydrogen-rich environment, which enhances the nuclear reaction chains
and energy release (Taam 1982, 1985).  All analyses (\cite{jos80};
\cite{aya82}; \cite{fus87b}; Taam, Woosley, \& Lamb 1996) have shown
that the helium burning in this mixed environment becomes thermally
stable when the local accretion rate is, by coincidence, super-Eddington
(see Bildsten 1998 for a review).  It is in this stable regime that we
carry out our calculations.

Finding the compositional makeup of the ashes from steady-state
hydrogen-rich accretion is complicated by the rp-process (\cite{wal81};
\cite{cha92}; \cite{van94}; \cite{sch98}), a sequence of rapid proton
captures onto nuclei that can produce proton-rich elements beyond iron.
The energy released is approximately known, however, so that we can
roughly determine the upper atmosphere's thermal profile for
steady-state burning.  Before discussing this problem, we first consider
the simpler case of pure helium accretion.  This is relevant to a number
of ultra-compact ($P_{\rm orb}< 50\minute$) binaries (4U~1626--67,
4U~1820--30, and 4U~1916--05, Nelson, Rappaport, \& Joss 1986;
potentially X1850--087, \cite{hom96}) and maybe the wide binary Cygnus
X-3, if it in fact contains a neutron star.

\subsection{Steady-State Pure Helium Burning}
\label{s:SteadyStateHeBurning}

There are a few neutron stars thought to be accreting nearly pure
helium.  For these stars, the first nuclear reaction is
\nuc{3\alpha\rightarrow{}^{12}C}.  We use the energy generation rate
from Fushiki \& Lamb (1987a) where $Q_{3\alpha}=7.274\MeV$ is the energy
released per formed carbon.  The radiative opacity $\kappa_{\rm rad}$ is
the sum of electron scattering (with high temperature and degeneracy
corrections from Paczy\'nski 1983) and free-free absorption.  We use the
conductive opacity, $\kappa_{\rm cond}=3\rho K({\rm
conductive})/4acT^3$, as given by Fushiki (1986).  The degenerate
electron pressure is calculated using an interpolation formula from
Paczy\'nski (1983).

The density at the location of helium burning decreases as the accretion
rate increases, which eventually allows the \C\alphagamma\Ox\ reaction
to compete with the $3\alpha$ reaction in consuming helium.  We use
Buchmann's (1996) reaction rate with the fitting formula that yields
$S=146 \keV{\rm\,barns}$ at 300 keV center-of-mass energy\footnote{There
is a typographical error in the fitting formula (\cite{buc96}, eq.\
[2]).  The terms $(1+p_2/T_9^{2/3})$ and $(1+p_6/T_9^{2/3})$ should be
squared (\protect\cite{buc97}).}.  This rate is roughly 1.7 times the
Caughlan \& Fowler (1988) \C\alphagamma\Ox\ rate, and so is more
consistent with nucleosynthetic constraints (\cite{wea93}).  Recent
measurements (\cite{azu94}) also support this {\em S}-factor value.  In
addition to \C\alphagamma\Ox, we also include the subsequent
\Ox\alphagamma\Ne, \Ne\alphagamma\Mg, and \Mg\alphagamma\Si\ reactions.
The reaction \Si\alphagamma$\rm^{32}S$ is unimportant for
$\mdot\le40\,\medd$, although for higher rates the reaction network
would have to include it as well.  For these reactions, we use rates
from Caughlan \& Fowler (1988) with the following adjustments: (1)
following Schmalbrock et al.~(1983), we set the uncertain factor
between 0 and 1 in the \Ne\alphagamma\Mg\ rate to 0.1; (2) we also set
the uncertain factor between 0 and 1 in the \Mg\alphagamma\Si\ rates to
0.1.  Screening is treated according to Salpeter \& Van Horn (1969).

We simultaneously integrate the steady-state versions of the continuity,
flux, and entropy equations (eqs.\ [\ref{e:cont}], [\ref{e:flux}],
[\ref{e:entropy}]) as a function of column depth.  Our integration
starts far beneath the photosphere, where the outward heat flux is set
by the energy released from nuclear reactions and the settling of the
material.  
Though the flux coming through the envelope is much less than the
accretion flux ($\approx GM\mdot/R$), the temperature just beneath the
photosphere is still near the value set by the accretion flux.  The flux
at the top of the atmosphere would simply be $F=(Q_{3\alpha}/12m_u)
\mdot=0.61\MeV(\mdot/m_u)$ if all the helium burned to carbon and there
was no energy released from gravitational settling.  However, we are
interested in the makeup of the products of helium burning; moreover,
the heat released from gravitational settling is non-negligible at high
accretion rates.
A hotter atmosphere has a greater scale height, and so at high accretion
rates more energy is released by lowering a parcel of fluid a distance
of order a scale height.
To integrate the fluid equations, we start with a trial value of the
flux and integrate equations (\ref{e:entropy}) and (\ref{e:flux}) to the
depth where all the helium is burned.  At this point, we compare the
flux to that coming from further gravitational settling of the ashes
(see \S~\ref{s:DeepOcean}) in deeper regions of the star, and
iteratively adjust the flux.

\begin{figure}[hbt]
\centering{\epsfig{file=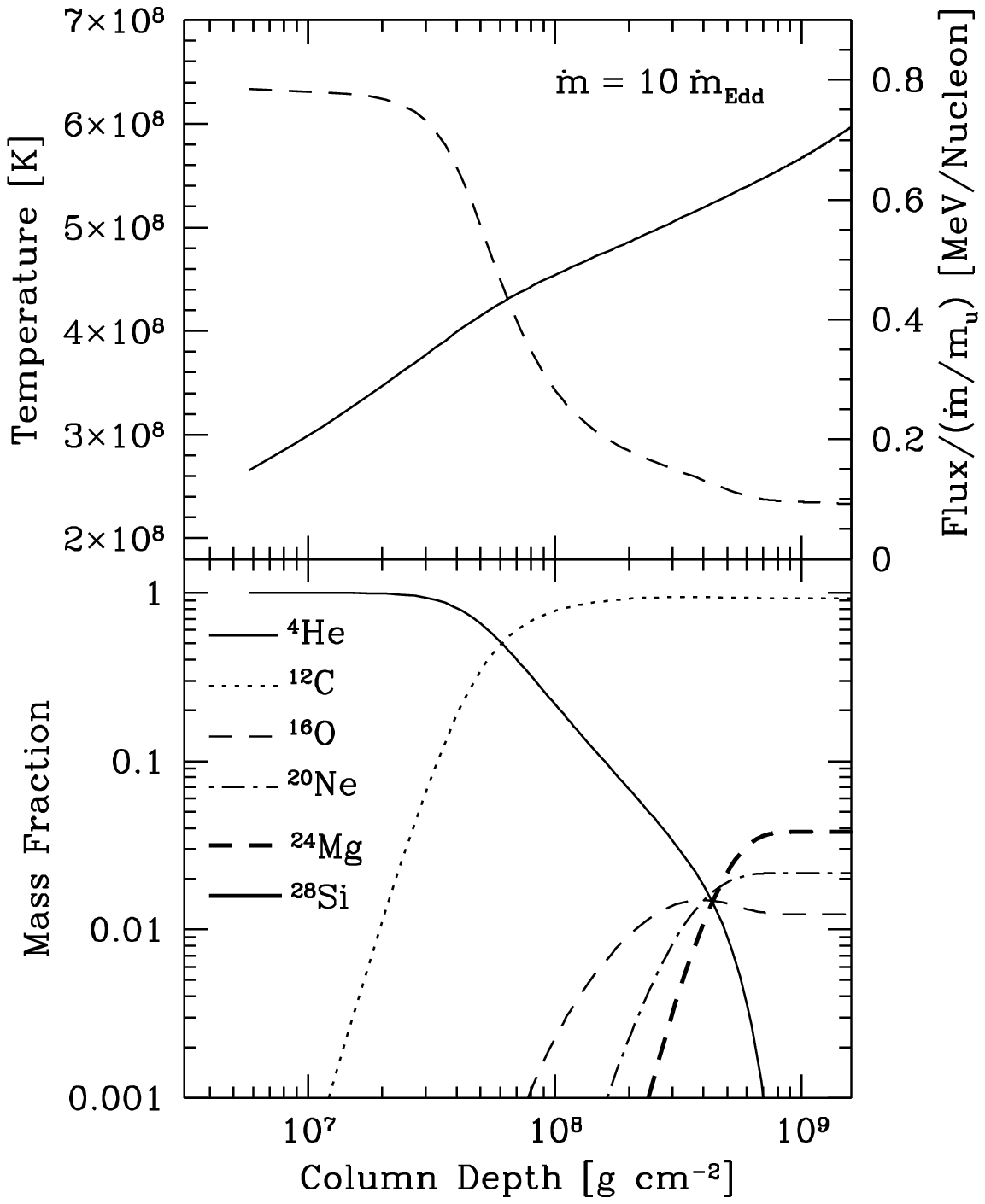,width=\hsize}}
\large\renewcommand{\baselinestretch}{0.7}\footnotesize
\caption{\protect{\footnotesize
Atmospheric temperature and composition for steady-state accretion and
burning of pure helium.  The neutron star is of mass $M=1.4\msun$ and
radius $R=10^6\cm$, and the accretion rate is
$\mdot=10.0\,\medd=7.5\ee{5}\GramPerSc\second^{-1}$.  The temperature is
shown by the solid line in the top panel, whereas the dashed line
displays the outward directed flux per accreted baryon.  The bottom
panel displays the mass fractions of the major elements present (\C,
\Ox, \Ne, \Mg, and \Si) for pure helium burning.
\label{f:helburn10}}}
\end{figure}

When burning in steady-state, the helium is consumed at the column depth
$y$ where its lifetime to the $3\alpha$ reaction is the time to reach
that depth, $\tfl\approx y/\mdot$.  For the values of $\mdot$ considered
here, burning occurs at a column depth $y\approx 10^8\GramPerSc$, which
is reached in a time $\tfl<10^3$ seconds.  Bildsten (1998)
provides a simple discussion of the physics of the thermal stability of
helium burning.  Linear perturbation calculations show that the helium
burning is thermally unstable when $T<5\times 10^8 \K$ in the burning
region (Bildsten 1995, 1998).  From our steady-state calculations and a
few time-dependent calculations, as in Bildsten (1995), we
find that pure helium burning is thermally unstable until
$\mdot\gtrsim10\,\medd$.

\begin{figure}[hbt]
\centering{\epsfig{file=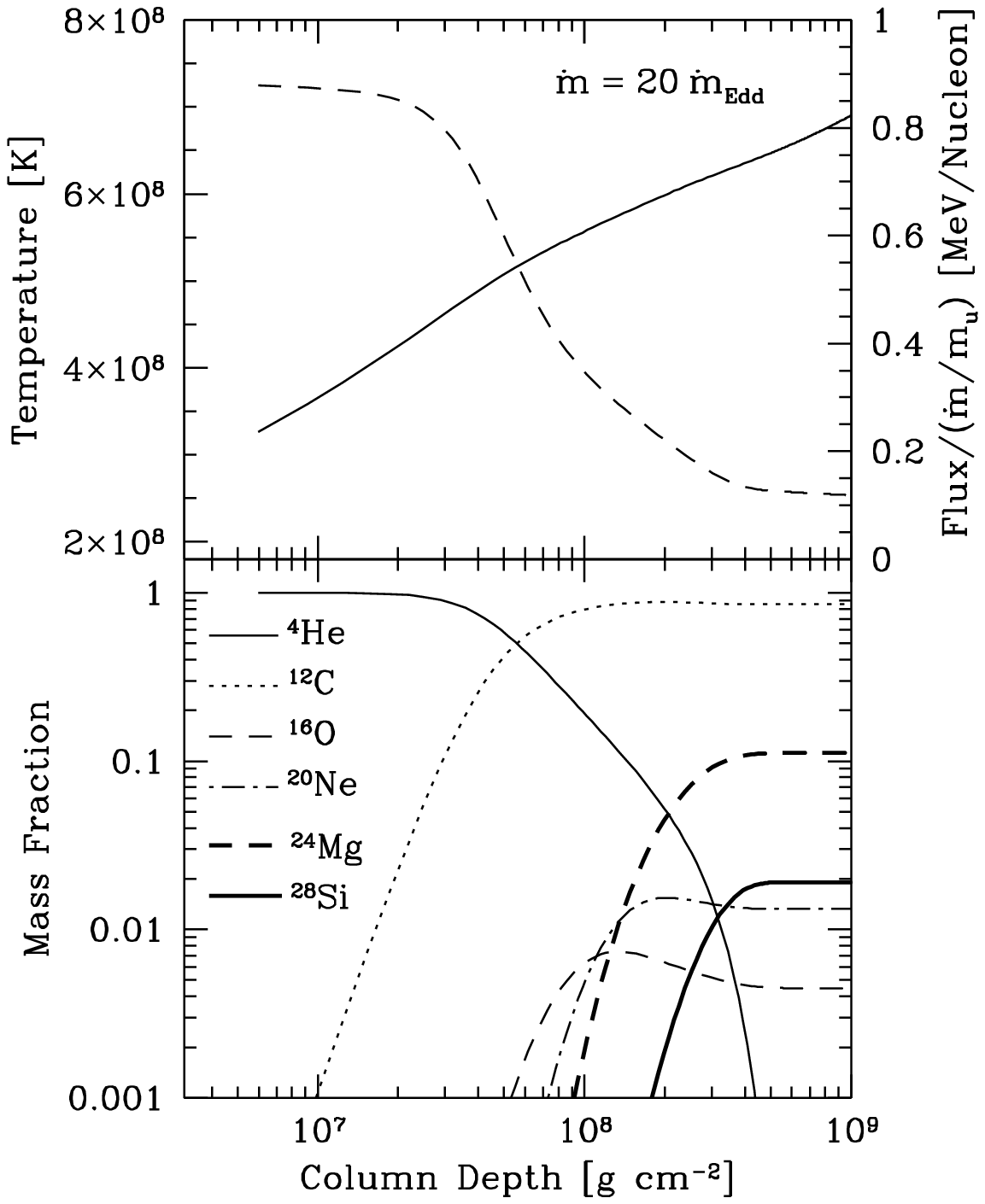,width=\hsize}}
\large\renewcommand{\baselinestretch}{0.7}\footnotesize
\caption{\protect{\footnotesize
Same as Fig.\ \protect\ref{f:helburn10}, but for an accretion rate
$\mdot=20.0\,\medd=1.5\ee{6}\GramPerSc\second^{-1}$.
\label{f:helburn20}}}
\end{figure}

\begin{figure}[hbt]
\centering{\epsfig{file=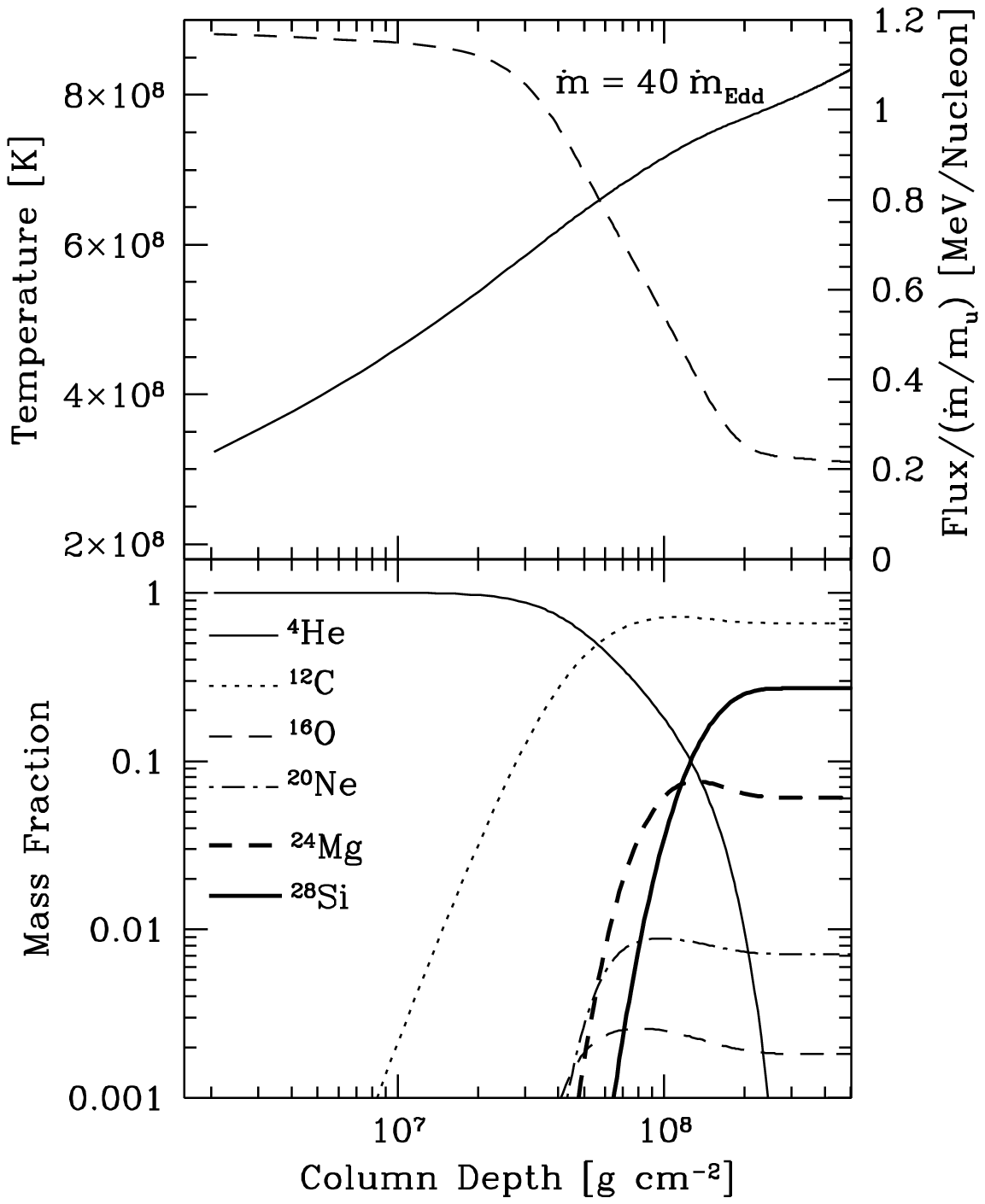,width=\hsize}}
\large\renewcommand{\baselinestretch}{0.7}\footnotesize
\caption{\protect{\footnotesize
Same as Fig.\ \protect\ref{f:helburn10}, but for an accretion rate
$\mdot=40.0\,\medd=3.0\ee{6}\GramPerSc\second^{-1}$.
\label{f:helburn40}}}
\end{figure}

Figures \ref{f:helburn10}, \ref{f:helburn20}, and \ref{f:helburn40}
display the steady-state thermal and compositional structure of the
outer layers of the neutron star accreting helium at rates of
$\mdot/\medd=10$, 20, and 40, respectively.  Each figure shows the
results for a different accretion rate.  The top panel displays the
temperature ({\em solid line\/}) and the flux per baryonic accretion
rate, $F/(\mdot/m_u)$ ({\em dashed line\/}).  The flux flowing through
the atmosphere increases with $\mdot$; this leads to increasingly hotter
temperatures and lower densities at the location of the burning and
subsequently more contributions from the $\alpha$ captures on carbon.
For each of Figures \ref{f:helburn10}, \ref{f:helburn20}, and
\ref{f:helburn40}, the composition as a function of column depth is
shown in the bottom panel.  As we show in Figure \ref{f:helburn40}, at
$\mdot=40\,\medd$ the final mixture is $\frac{2}{3}$ carbon and $\frac{1}{3}$ silicon.
At lower accretion rates, the dominant nucleus made from helium burning
is carbon.  The outward-directed flux at the bottom of the atmosphere
comes from the gravitational settling of matter beneath the helium
burning region.  This flux is determined by continuing the integration
of equations (\ref{e:entropy}) and (\ref{e:flux}) to very deep parts of
the neutron star, as described in section \ref{s:DeepOcean}.

\subsection{Steady-State Burning of Accreting Hydrogen and Helium} 
\label{s:SteadyStateHHeBurning}

At these high accretion rates, the hydrogen cannot burn rapidly enough
to be depleted prior to helium ignition.  As a result, hydrogen burning
proceeds via the rp-process, which is accelerated by the proliferation
of ``seed'' nuclei from helium burning.  As a first guess, we presume
that the hydrogen burns completely where the helium ignites and neglect
calculating the composition of the ashes.  These atmospheres are much
hotter than the pure helium burning case (for the same $\mdot$) due to
the larger flux flowing through the envelope; this leads to thermally
stable helium burning once $\mdot\gtrsim\medd$ (\cite{bil98a}).

We presume that the flux flowing through the atmosphere is that released
from burning all of the hydrogen and helium to carbon,
\begin{equation} 
   F=\frac{\mdot}{m_u}
   	\left[X\left(\frac{Q_H}{4}+\frac{Q_{3\alpha}}{12}\right) 
   	+Y\frac{Q_{3\alpha}}{12}\right],
\label{e:fluxhyd}
\end{equation}
where $Q_H=28\MeV$ is the energy release per formed helium.  This
approximation yields lower limits to the true temperature at the
location of helium burning in the presence of the more complex
rp-process for two reasons.  First, we presume that the reactions
proceed only to carbon, although the residual nuclei are likely to be
much heavier.  In addition, Taam et al.~(1996) have found (for high
accretion rate type I bursts) that not all of the hydrogen is consumed
at the location where the helium is depleted, in which case there is
further energy release from electron capture on protons at larger depths
in the atmosphere (for a recent discussion, see \cite{bil98b}).  {\em
Therefore, our thermal profiles for depths greater than where helium
burns should be considered as lower limits.}

\begin{figure}[hbt]
\centering{\epsfig{file=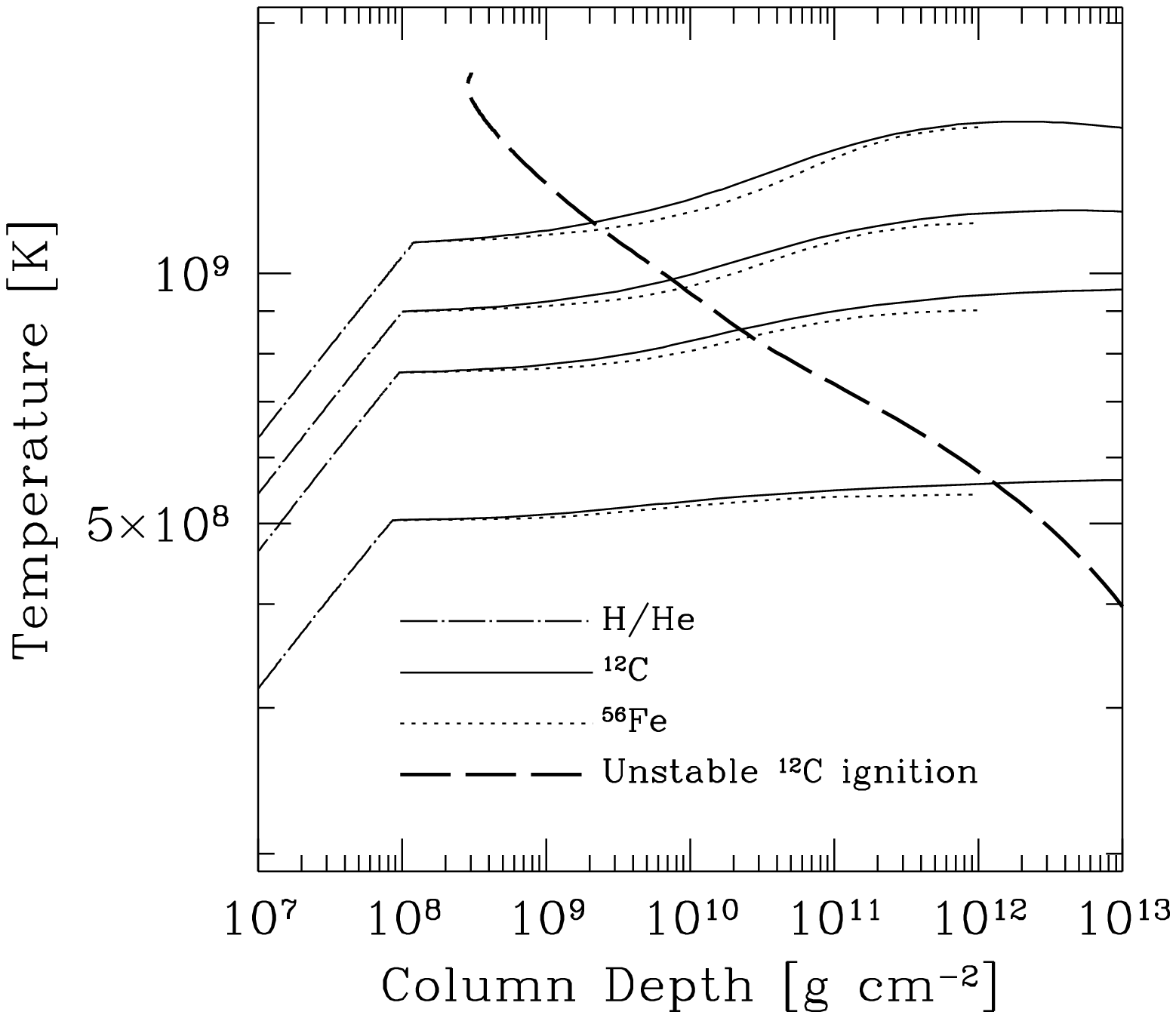,width=\hsize}}
\large\renewcommand{\baselinestretch}{0.7}\footnotesize
\caption{\protect{\footnotesize
Atmospheric temperature profiles for hydrogen-rich accretion ($X=0.7,\
Y=0.3$) at the super-Eddington accretion rates (from bottom to top)
$\mdot/\medd=1$, 5, 10, and 20.  In the outer atmosphere, the flux is
set by hydrogen/helium burning ({\em dot-dashed curves\/}).  Where the
helium is half-depleted, we join these curves onto the settling
solutions (no burning) for an ocean of either carbon ({\em solid
curves\/}) or iron ({\em dotted curves\/}).  The slopes are
discontinuous at this point because the flux released from settling and
compression alone is much less than the burning flux.  If any carbon is
present, it will ignite where $d\enuc/dT=d\ecool/dT$ (see \S~7.1; {\em
heavy dashed line\/}).  The iron settling solutions end at a lower
column depth than the carbon settling solutions because of
crystallization.  At column depths greater than $\sim10^{9}\GramPerSc$,
the density is approximately given by eq.\ (\protect\ref{e:rho(y)}).
\label{f:Hsettle}}}
\end{figure}

For accretion rates $\mdot/\medd=1.0$, 5.0, 10.0, and 20.0 and a helium
fraction $Y=0.3$ (Fig.\ \ref{f:Hsettle}), we integrate equation
(\ref{e:flux}), with the flux given by equation (\ref{e:fluxhyd}),
downwards into the atmosphere until one-half of the incident helium has
burned.  At this point, we presume that all the hydrogen has burned as
well, in which case the flux originating from deeper points is entirely
caused by gravitational settling.  For definiteness, we took the
composition to be either carbon ({\em solid curve\/}) or iron ({\em
dotted curve\/}).  The helium ignition depth (clearly denoted by the
discontinuity in slope) is nearly independent of the accretion rate,
yielding times to helium ignition of roughly $0.4\,(\medd/\mdot)\hour$.
At the highest accretion rate shown, the infalling helium ignites only
80~s after arrival onto the neutron star.  The helium burning is
stable once $\mdot\gtrsim\medd$; the cooling rate at the temperature
($5\times10^8\K$) in the helium burning region at that accretion rate is
fast enough to stabilize the burning.

\section{Gravitational Compression of the Ashes in the Deep Ocean}
\label{s:DeepOcean}

We now discuss the thermal state of the degenerate ocean underneath the
hydrogen/helium burning region.  Upon completion of stable
hydrogen/helium burning, the gas is hotter than $5\ee{8}\K $ and denser
than $(\mbox{2--5}) \ee{5}\GramPerCc$.  The temperature and
nuclear evolution of this matter under further compression depends on
both the local accretion rate and the composition.

First consider steady-state settling solutions in the absence of further
nuclear burning.  In this case the continuity equation
(eq.~[\ref{e:cont}]) is identically satisfied, so our equations are
\begin{mathletters}
\begin{equation} \label{e:st_T}
   \DDy{T} = \frac{F}{\rho K}
\end{equation}
and
\begin{equation}\label{e:st_ent}
   \DDy{F} = c_{\!P}\DDt{T}-\frac{c_{\!P}T\mdot}{y}\delab+\enu,
\end{equation}
\end{mathletters}
where $\enu$ is the neutrino cooling rates from Schinder et al.~(1987).
From these equations we define the characteristic timescale for neutrino
cooling, $\tnu\equiv c_{\!P} T/\enu$.  As long as $T\lesssim 10^9\K$, this
time is typically longer than either the time for heat to diffuse across
a scale height, $\tth \equiv c_{\!P} y^2/K\rho$, or the time for the
temperature to increase under adiabatic gravitational compression, $\tgh
= y/\mdot\delab$.  As a result, the thermal profile of the ocean is
determined by the competition between $\tth$ and $\tgh$; comparing these
timescales defines a {\em local\/} critical accretion rate as a function
of depth,
\begin{equation}
   \mcrit \equiv \frac{\rho K}{c_{\!P} y}\delab{}^{-1}.
\end{equation}
The quantity $\mdot/\mcrit$ characterizes the thermal profile of the
ocean.  For $\mdot\gg\mcrit$, thermal diffusion is negligible in the
time required for compression to occur, so that the flow approaches
adiabatic compression: $\nabla = \delab$.  In the opposite limit of low
accretion rates ($\mdot\ll\mcrit$) there is plenty of time for the heat
to diffuse and the flow is far from adiabatic.  We can then rewrite
equation (\ref{e:st_ent}) as
\begin{equation}\label{e:mdotllmcrit}
   \DDy{F} = -\frac{c_{\!P}T\mdot}{y}\delab.
\end{equation}
In this case $\nabla$ is self-consistently small.  We now show that, in
the deep ocean, $\mcrit$ exceeds the Eddington limit by nearly two
orders of magnitude.

\subsection{ A Simple Example of the Critical Accretion Rate in the Ionic
Ocean} \label{s:IonicOcean}

We first duplicate the original case discussed by Bildsten \& Cutler
(1995).  Consider a deep ocean where the degenerate electrons are
extremely relativistic ($\rho \gtrsim 10^7\GramPerCc$) and the ions
(presumed for simplicity to be of a single species of charge $Ze$ and
mass $Am_u$) are not crystallized.  The equation of state is then
roughly
\begin{equation} \label{e:EquationOfState}
   P \approx \frac{1}{4}n_e m_e c^2 x +  n_I\kB T,
\end{equation}
where $n_e$ ($n_I$) is the number density of electrons (ions), $m_e$ is
the electron mass, and
$x\equiv\pF/m_ec=1.008(Z\rho/A10^6\GramPerCc)^{1/3}$ is the relativity
parameter ($\pF$ is the electron Fermi momentum).  The pressure scale
height is then
\begin{eqnarray} \label{e:ScaleHeight}
   H &\approx& R \left(\frac{Z}{4A}\right)
   \left(\frac{m_ec^2}{GMm_u/R}\right) x \nonumber\\
   &=& 265 \left(\frac{2Z}{A}\right)^{4/3} 
   \left(\frac{\rho}{10^6\GramPerCc}\right)^{1/3}\cm,
\end{eqnarray}
and the density as a function of column depth is
\begin{equation}\label{e:rho(y)}
   \rho \approx 7.6\ee{6} \left(\frac{A}{Z}\right)
   \left(\frac{y}{10^{10}\GramPerSc}\right)^{3/4}\GramPerCc.
\end{equation}
The contribution from electron scattering to the conductivity is
approximated by the Wiedemann-Franz law,
\begin{equation}\label{e:Wiedemann-Franz}
   K(\mbox{\rm electron scattering}) = 
      \frac{\pi^2}{3}\frac{\kB^2 T}{m_e\sqrt{1+x^2}} n_e \nu^{-1}, 
\end{equation}
where $\nu=\nu_{\rm ee}+\nu_{\rm ei}$ is the sum of electron-electron
and electron-ion collision frequencies.  We use a convenient fitting
formulae (\cite{tim92a}) for $\nu_{\rm ee}$ and the formulae of Yakovlev
\& Urpin (1980) for $\nu_{\rm ei}$.

The entropy per unit mass is (\cite{lan80})
\begin{eqnarray}\label{e:s}
   s &=& s_{\rm ion} + s_{\rm elec.} \nonumber\\
   &\approx& \frac{\kB }{A m_u} \left[\ln\left(\frac{T^{3/2}}{\rho}\right) +
      \pi^2 Z \frac{\kB T}{m_ec^2 x}\right],
\end{eqnarray}
so that $\delab\approx\frac{1}{2}$, and the specific heat at constant
pressure (see, e.g., \cite{cox68}, \S~9.13) is
\begin{eqnarray} \label{e:c_P}
   c_{\!P} &=& T\left(\frac{\partial s}{\partial T}\right)_\rho +
      \frac{P}{\rho T} 
      \left(\frac{\partial \ln P}{\partial \ln T}\right)_\rho^2 
      \left(\frac{\partial \ln P}{\partial \ln\rho}\right)_T^{-1}
	\nonumber\\
   &\approx& \frac{\kB }{A m_u}\left[\frac{3}{2} + 
      {\cal O}\left(\frac{\kB T}{m_ec^2x}\right)\right].
\end{eqnarray}
The critical accretion rate for this ocean is then
\begin{eqnarray}
   \mcrit &\approx&  10^8 \localrate\left(\frac{T}{5\ee{8}\K}\right)
      \left(\frac{A}{2Z}\right)^2 \nonumber\\
	&\gg& \medd. 
\end{eqnarray}
For $\mdot\ll \mcrit$ and a nearly isothermal ocean, we find the flux in
the deep ocean by integrating equation (\ref{e:mdotllmcrit}), which
yields (\cite{bil95b})
\begin{equation}\label{e:analyticF}
   F = \frac{3}{4}\frac{\dot{m}\kB T}{A m_u}
      \ln\left(\frac{y_0}{y}\right) + F_0,
\end{equation}
where $F_0$ is the flux at a fiducial column depth $y_0 > y$.
This solution serves as a useful guide to understanding our more
accurate solutions, as we will consider accretion rates in excess of
$\medd$, but well below $\mcrit$.

\begin{figure}[hbt]
\centering{\epsfig{file=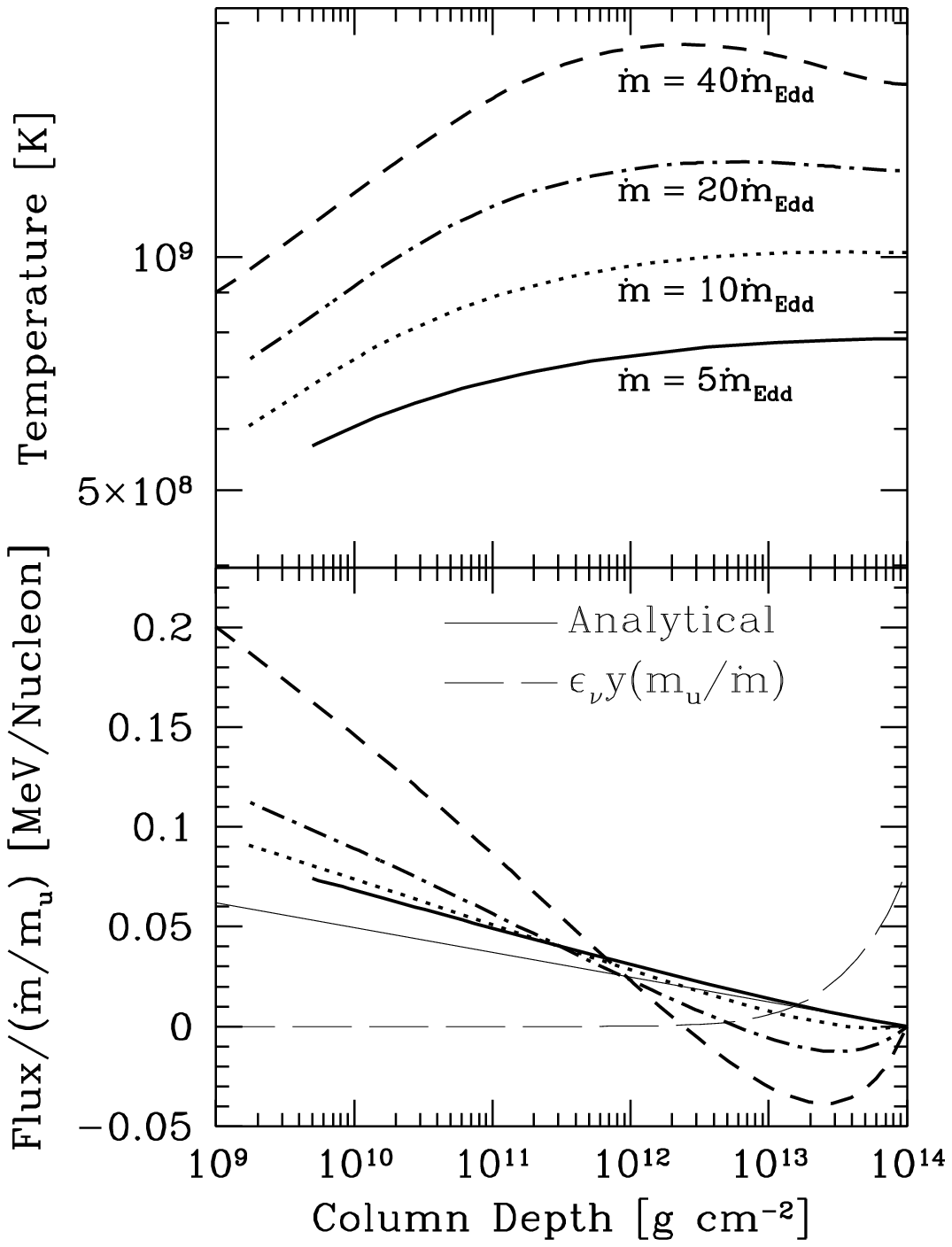,height=280pt}}
\large\renewcommand{\baselinestretch}{0.7}\footnotesize
\caption{\protect{\footnotesize
Temperature (top panel) and flux (bottom panel) as a function of column
depth for a carbon ocean formed by the steady state burning of accreted
pure helium.  The density as a function of column depth is approximately
given by eq.\ (\protect\ref{e:rho(y)}).  The accretion rates are
$\mdot/\medd=5.0$ ({\em solid line\/}), 10.0 ({\em dotted line\/}), 20.0
({\em dot-dashed line\/}), and 40.0 ({\em short dashed line\/}).  For
comparison, the flux predicted by eq.\ (\protect\ref{e:analyticF}) is
depicted ({\em light solid line\/}) in the bottom panel for a
temperature of $10^9\K$.  We also show the approximate integrated
neutrino emissivity, $\enu y(m_u/\mdot)$ ({\em light dashed line\/}), in
units of MeV per nucleon for the $\mdot=40.0\,\medd$ solution.  As is
evident from this solution, when neutrino cooling is important, the
thermal profile is sensitive to the location of the bottom boundary.
This fact motivates our study of how the ocean and crust are thermally
coupled to the core.
\label{f:HeSettle}}}
\end{figure}

\subsection{Accurate Solutions for the Compressed Ashes}\label{s:solutions}

For the case $\medd<\mdot\ll\mcrit$, so that $\nabla\ll\delab$, we now
numerically solve equations (\ref{e:st_T}) and (\ref{e:st_ent}) for
the settling of matter in the deep ocean.  We include neutrino cooling
and include nonideal
(ion-ion electronic) contributions to the equation of state.  For the
electrons' contribution to the specific heat, we use the interpolation
formula of Paczy\'nski (1983).  The ions are in a liquid state
($1<\Gamma\lesssim 170$), where
\begin{eqnarray}\label{e:Gamma}
   \Gamma &\equiv& \frac{Z^2 e^2}{\kB T}
      \left(\frac{4\pi n_I}{3}\right)^{1/3}\nonumber\\
   &\approx& 1.1 \left(\frac{10^8\K}{T}\right)
      \left(\frac{\rho}{10^8\GramPerCc}\right)^{1/3} \frac{Z^2}{A^{1/3}}
\end{eqnarray}
is the ion parameter for a single-component fluid.  We use the
parameterization of Farouki \& Hamaguchi (1993) to account for the
electrostatic contribution to the ionic energy.  This parameterization,
although defined only for $\Gamma\ge 1$, is, for $0.3<\Gamma<1.0$,
within ten percent of the considerably more cumbersome parameterization
of Hansen (1973), which is valid for $0<\Gamma\lesssim 160$.  For the
conditions of interest, $\Gamma$ is always greater than 0.4; we
therefore use the parameterization of Farouki \& Hamaguchi (1993) for all
$\Gamma$ so that the specific heats are continuous.

The numerical integration of equations (\ref{e:st_T}) and
(\ref{e:st_ent}) requires two boundary conditions.  The temperature at
the top of the ocean is determined by the solution in the hydrogen/helium
burning region (\S~\ref{s:He_burn}), whereas the flux leaving the top of
the ocean (and entering the atmosphere) depends on the gravitational
settling in deeper regions.  We thus must specify either the temperature
or flux at the bottom of the deep ocean.  For the initial calculation,
we presume that the flux at the bottom of the ocean is zero (but see
\S~\ref{s:core}).  Figure \ref{f:HeSettle} depicts the resulting
temperature ({\em top panel\/}) and flux ({\em bottom panel\/}) profiles
for the ashes 
of pure helium burning under accretion rates $\mdot/\medd=5.0$ ({\em
solid line\/}), 10.0 ({\em dotted line\/}), 20.0 ({\em dot-dashed
line\/}), and 40.0 ({\em short dashed line\/}).  As was shown in
\S~\ref{s:He_burn}, for these accretion rates, the ashes will be very
nearly pure carbon.  For accretion rates below $40.0\,\medd$, the flux
roughly obeys $F\propto\ln y$ (eq.\ [\ref{e:analyticF}]); this analytic
solution is plotted ({\em light solid line\/}) for $T=10^9\K$ in the
bottom panel of Figure \ref{f:HeSettle}.  The slopes are steeper than
that predicted by equation (\ref{e:analyticF}) because of ion-ion
interactions, which increase both the specific heat and the value of
$\delab$.  Cooling by neutrino emission ({\em light dashed line\/})
clearly affects the highest accretion rate ($40.0\,\medd$) solution for
$y>10^{12}\GramPerSc$ and $T>10^9\K$.  Note that if the neutrino cooling
term is significant in equation (\ref{e:st_ent}), then the thermal
profile is sensitive to the location of, and the flux at, the bottom
boundary.  Properly connecting the thermal equations of the ocean to
those of the crust and core is the topic of the next subsection.  We
discuss, in section \ref{s:ignition}, the stability of carbon burning in
oceans this hot.

\begin{figure}[hbt]
\centering{\epsfig{file=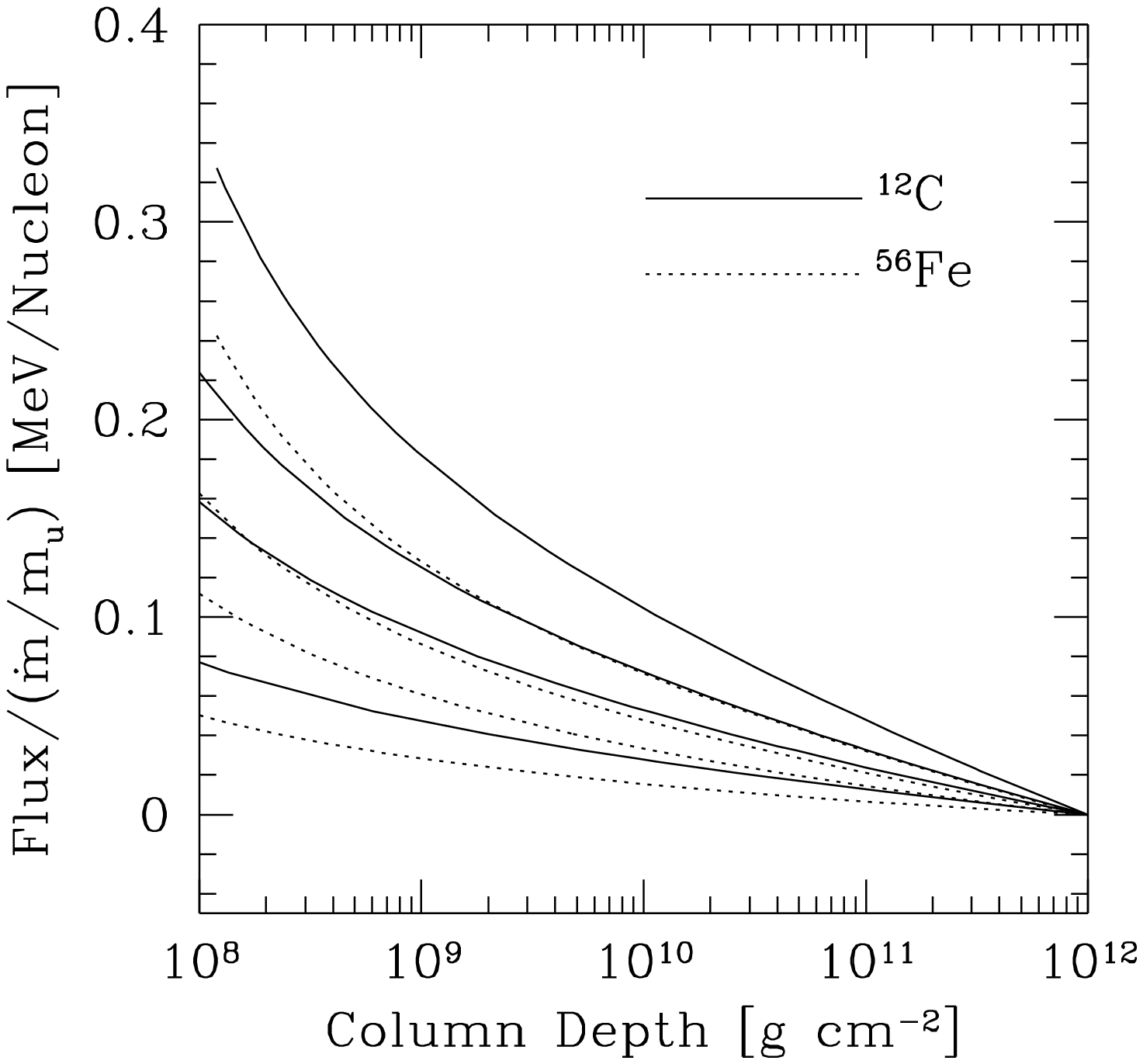,width=\hsize}}
\large\renewcommand{\baselinestretch}{0.7}\footnotesize
\caption{\protect{\footnotesize
Flux per accreted baryon as a function of column depth, for accretion of
hydrogen-rich matter.  We show both a carbon ocean ({\em solid lines\/})
and an iron ocean ({\em dotted lines\/}).  In order of increasing flux
at the top ($y=10^9\GramPerSc$), the curves are for accretion rates
$\mdot/\medd=1.0$, 5.0, 10.0, and 20.0.  We use the boundary condition
$F(y=10^{12}\GramPerSc)=0$ in both cases.  At column depths greater than
$\sim10^{9}\GramPerSc$, the density is approximately given by eq.\
(\protect\ref{e:rho(y)}).
\label{f:HSettleFlux}}}
\end{figure}

For completeness, we also show in Figure \ref{f:HSettleFlux} the flux
per accreted baryon if the star is accreting a hydrogen and helium mix.
Because the products of H/He burning are uncertain in this case (see
\S~\ref{s:He_burn}), we include a carbon ocean ({\em solid lines\/}) and
an iron ocean ({\em dotted lines\/}), both accreting at rates
$\mdot/\medd=1.0$, 5.0, 10.0, and 20.0.  For comparison, we have
integrated both cases only to a column depth of $10^{12}\GramPerSc$,
corresponding roughly to the depth where iron crystallizes.  As a
consequence, neutrino cooling does not affect the thermal profile, so
that the flux is always positive for $y>10^{12}\GramPerSc$.  The
difference in the flux for nearly identical thermal profiles (see Fig.\
\ref{f:Hsettle}) is due to the differences in specific heat between an
iron and a carbon ocean.  The ideal gas specific heat is proportional to
$1/A$ (eq.\ [\ref{e:c_P}]).  However, the stronger ion-ion interactions
of the iron ocean partially offset this reduction in $c_{\!P}$ and also
increase the value of $\delab$.  The net result is that for a carbon
ocean, the slope of $F/(\mdot/m_u)$ is about 1.5 times greater than the
slope for an iron ocean.

\subsection{Heating and Cooling of the Crust and Core}
\label{s:core}

The steady-state thermal profiles found in the previous section assumed
a zero-flux bottom boundary condition, or equivalently, an
asymptotically isothermal temperature profile.  If we were to assume a
core temperature equal to this temperature, then a substantial neutrino
flux would be generated.  This alerts us to the facts that neutrino
cooling in the crust is important and that neutrino emission from the
core will have an effect on the crust and ocean even for
$\mdot\lesssim\medd$.  The boundary conditions at the bottom of the
ocean must account for this.

For a core composed of superfluid neutrons and protons, the neutrino
luminosity is dominated by crust bremsstrahlung.  For illustrative
purposes, we use the formula (\cite{shi89})
\begin{equation}
   L_\nu \approx 4\ee{31}\erg\second^{-1}
   \left(\frac{T_c}{10^8\K}\right)^6, 
\end{equation}
which demands a comparable heating of the core by conduction.  There are
two heat sources in the deep crust.  First, the original crust of an
accreting neutron star will be replaced in less than $10^7\yr$ when
$\dot{M} > 10^{-8}\,\msun\yr^{-1}$.  As accretion pushes the crust
deeper, the continuous compression forces the crust matter to endure a
series of electron captures, neutron emissions, and pycnonuclear
reactions (\cite{hae90}).  The heat deposited into the neutron star from
these reactions is $\approx 1\MeV/{\rm nucleon}$ and occurs at 
column depths $\gtrsim 10^{14}\GramPerSc$.  There is also heat generated
directly from compression, which can at best generate about
$0.5\MeV/{\rm nucleon}$.  If the total electron capture flux, integrated
over $4\pi$ steradians, is inward-directed so that it diffusively heats
the core, then in steady state this flux is equal to $L_\nu$, which
implies an equilibrium core temperature
\begin{equation}\label{e:SteadyStateTc}
    T_c \approx  5\ee{8}\K
   \left(\frac{\dot{M}}{10^{-8}\,\msun\yr^{-1}}\right)^{1/6}.
\end{equation}
However, an inverted temperature gradient is needed to carry the flux,
so that a core in steady-state implies a temperature maximum somewhere
in the ocean or crust.
Throughout this subsection, we shall assume spherically symmetrical
accretion.

In order to estimate the thermal profile in the crust, we integrate
equation (\ref{e:st_T}) for constant flux.  In the Wiedemann-Franz
approximation (eq.\ [\ref{e:Wiedemann-Franz}]), the thermal conductivity
$K$ is determined by the collision frequency of electrons with other
electrons, with ions, and---where the ions are crystallized---with
phonons and lattice impurities.  Where the ions are crystallized and the
temperature is greater than the Debye temperature,
\begin{equation}
   \Theta = 1.76\ee{8}\K \left(\frac{2Z}{A}\right)
	\left(\frac{\rho}{10^{10}\GramPerCc}\right)^{1/2},
\end{equation}
we use the electron-phonon collision frequency obtained with a
relaxation approximation (\cite{yak80}), $\nu_{\rm ph} \approx 13 e^2
\kB T/(\hbar^2c)$.  Although this formula is valid only when
$T\gtrsim\Theta$,
the steady-state temperature (eq.\ [\ref{e:SteadyStateTc}]) is so high
that neglecting the $T<\Theta$ correction does not contribute a
significant error (less than a factor of 3) to $K$ for $\rho \lesssim
10^{12}\GramPerCc$.  We then analytically integrate equation
(\ref{e:st_T}) inward in column depth from $y_0$ to $y$ to find the
flux,
\begin{eqnarray}\label{e:FluxInCrust}
   \lefteqn{F(y) \approx }\\ \nonumber
       && 7.9\ee{9} \left(\frac{A}{Z}\right) y_0{}^{1/4}
	\frac{T(y)-T(y_0)}{1-(y_0/y)^{1/4}}\erg\second^{-1}\cm^{-2},
\end{eqnarray}
where $T_0$ and $F_0$ are the temperature and flux, respectively, at
$y=y_0$.

\begin{figure}[hbt]
\centering{\epsfig{file=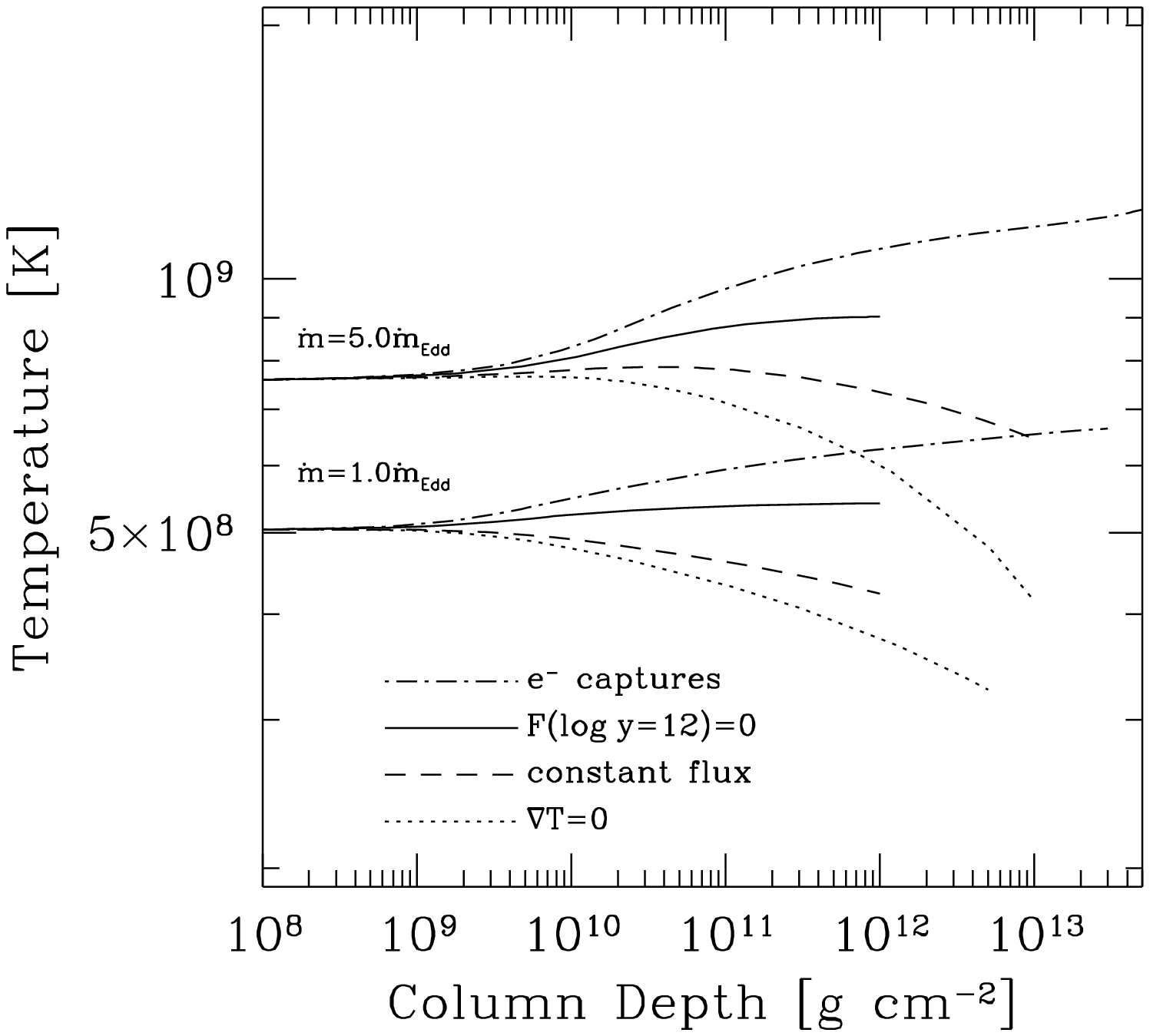,width=\hsize}}
\large\renewcommand{\baselinestretch}{0.7}\footnotesize
\caption{\protect{\footnotesize
Thermal profiles in the ocean for various models of the crust/core.  The
star is accreting hydrogen-rich ($X=0.7,\ Y=0.3$) matter at rates
$\mdot/\medd = 1.0$ and 5.0.  For each accretion rate, we compute three
different models: (1) a constant-flux crust with heat generation from
electron captures ({\em dot-dashed lines\/}); (2) an isothermal crust
with no heat sources other than compression ({\em dotted lines\/}); and
(3) a constant-flux crust with no heat sources ({\em dashed lines\/}).
For comparison, we also display the settling solutions from Figure
\protect\ref{f:Hsettle} ({\em solid lines\/}), for which
$F(y=10^{12}\GramPerSc)=0$.  At column depths greater than
$\sim10^{9}\GramPerSc$, the density is approximately given by eq.\
(\protect\ref{e:rho(y)}).
\label{f:core_models}}}
\end{figure}

A detailed solution of the thermal and compositional evolution of the
inner crust and core of an accreting neutron star is beyond the scope of
this paper.  However, we can use our results for the deep ocean to
construct models, which will serve as a check on our choice of
bottom boundary conditions in the integration of the fluid equations.
We show the oceanic thermal profiles from these models, for accretion
rates $1.0\,\medd$ and $5.0\,\medd$ (hydrogen-rich accretion), in Figure
\ref{f:core_models}, along with the iron settling solutions from Figure
\ref{f:Hsettle} ({\em solid lines\/}).

In the first model ({\em dot-dashed lines\/}), we assume that the energy
sources in the crust are concentrated at a single point $y_w$.  A
fraction $f$ of the generated flux is directed outward, while a fraction
$(1-f)$ is directed inward.  Because the nuclear energy released
(about 1~MeV/nucleon) is much greater than the energy released
from compression, we assume that the crust flux is constant away from
$y_w$ and integrate equation (\ref{e:st_T}) to obtain the temperature in
the crust as a function of column depth.  We used the hydrogen and
helium burning calculations described in \S~\ref{s:He_burn} with the
assumption that the accreted material is processed to iron where the
hydrogen burns in steady state.  Using this assumption to fix the
temperature at the top of the iron ocean, we then numerically integrate
the fluid equations to the depth where iron crystallizes (from eqs.\
[\ref{e:Gamma}] and [\ref{e:EquationOfState}] this is $y_\Gamma\sim
5\ee{13}\GramPerSc$).  This solution implies a flux $F_{\rm\!out}$
flowing outwards from the crust.  Using $F_{\rm\!out}=f\mdot(1\MeV/m_u)$
to fix $f$, we then solve equation (\ref{e:FluxInCrust}) to find the
temperature at $y_w$.  From there, we again solve equation
(\ref{e:FluxInCrust}) with an inward-directed flux
$F_{\rm\!in}=-(1-f)\mdot(1\MeV/m_u)$ to find the core temperature $T_c$.
We then adjust $f$ until the radiative luminosity entering the core
balances the neutrino luminosity.  Because the thermal conductivity
increases with depth, the flattest temperature gradient is realized by
placing the heat source as deep as possible.  We set this point at a
column depth $y_w = 10^{17}\GramPerSc$.  As an example, for an Eddington
accretion rate, the maximum temperature in the crust is $T(y_w) =
7.0\ee{8}\K$, and the core temperature is $T_c = 5.3\ee{8}\K$ with a
corresponding neutrino luminosity $L_\nu = 8.8\ee{35}\erg\second^{-1}$.
About 2\% of the heat generated in the crust diffuses upward into the
ocean; although not inconsequential, the change in temperature does not
significantly change the results of our integration of the settling
solutions.  Including the contributions of {\em Umklapp\/} and impurity
scattering to the conductivity only steepens the temperature gradient
between $y_w$ and the core, so that $T(y_w)$ would be even hotter.

In the second and third models (which are somewhat unrealistic), we
suppose the crust to have no heat sources other than compression.
Because of the high conductivity, $\nabla$ is much less than $\delab$
throughout the crust (see \S~\ref{s:IonicOcean}); nevertheless, the core
neutrino luminosity demands an inward-directed flux.  To estimate the
effect of compressional heating in the crust on the equilibrium core
temperature and the oceanic thermal profiles, we performed two
integrations, one with a constant flux (integration of eq.\
[\ref{e:st_T}]; {\em dotted lines\/}) and one with a constant
temperature (integration of eq.\ [\ref{e:mdotllmcrit}]; {\em dashed
lines\/}) crust.  The constant-flux crust is completely devoid of any
heat sources, while in the isothermal crust model, heat is generated via
compression.  For both cases, we numerically integrate over the ocean,
and then integrate analytically either equation (\ref{e:st_T}) or
equation (\ref{e:mdotllmcrit}) to estimate the heat flux into, and the
temperature of, the core.  We again balance the flux directed into the
core with the neutrino luminosity, and adjust the flux at the top of the
ocean until the two match.  For a constant-flux integration at
$\mdot=\medd$, the core temperature and neutrino luminosity are
$T_c=3.0\ee{8}\K$ and $L_\nu=2.3\ee{34}\erg\second^{-1}$, respectively.
The isothermal crust model produces a slightly higher temperature and
luminosity: $T_c=3.3\ee{8}\K$ and $L_\nu=4.8\ee{34}\erg\second^{-1}$.

As a check of our model assumptions, we compute $\nabla/\delab$ at the
top of the crust (where this ratio will be greatest).  Because of the
strong sensitivity of $L_\nu$ to temperature, the core temperature is
roughly the same in all cases.  We also compute the ratio of the
settling flux (from eq.\ [\ref{e:mdotllmcrit}]) generated in the crust
to that entering the core (in the model where the flux is assumed
constant, we calculate the settling flux that would have been
generated).  For sub-Eddington accretion rates, either approximation
will serve---the gradient is very sub-adiabatic and the settling flux in
the crust is small compared to that required by the neutrino luminosity.
For higher accretion rates, $\nabla/\delab$ is no longer negligible, so
that the constant-flux approximation is better, as most of the settling
flux is generated in the upper ocean (see eq.\ [\ref{e:mdotllmcrit}]).
This regime is unlikely to be important, as the accreted matter will
have spread around the stellar surface at this depth, so that the local
accretion rate is not more than a few times Eddington.

Previous authors (\cite{aya82}; \cite{fuj84}; \cite{mir90}) also
considered the total energy balance of the neutron star core and its
influence on the outer crust, albeit for much lower accretion rates.
Because of the rapid compression of the star's outer layers and the
reactions in the deep crust, the outer layers of the neutron star do not
approach an isothermal temperature profile.  In contrast, Ayasli \& Joss
(1982) considered densities greater than $10^8\GramPerCc$ to be part of
an isothermal core.  Their calculation of neutrino luminosities differs
from ours in that they include general relativistic corrections to the
core temperature (their standard model was for a star of mass
$1.41\,\msun$ and radius $6.57\km$, so that the relativistic corrections
are quite nonnegligible).  The corrections imply a higher neutrino
luminosity, as measured by an observer at infinity, for a given core
temperature.  Fujimoto et al.~(1984) considered the flow of heat into
the core (as well as different core models), but again, their typical
crust temperatures were less than $10^8\K$.  Like Ayasli \& Joss (1982),
their models had heat sources only in the outer envelope.
Non-equilibrium reactions in the crust were first incorporated into
accreting neutron star models by Miralda-Escud\'e et al.~(1990).  They
did not consider local accretion rates greater than roughly 0.01 times
the Eddington rate (using their radii), so that the crust temperatures
were typically never greater than $10^8\K$.

  In conclusion, the physical processes deep in the crust do not
appreciably affect the thermal profiles in the upper ocean, which is
dominated by compressional heating and nuclear processing of hydrogen
and helium.  It is important therefore to treat the upper atmosphere
correctly, as the thermal profiles of the ocean depend completely on
the physics in that region.

\section{Ohmic Diffusion in the Deep Ocean and Crust}
\label{s:OhmicDecay}

Present uncertainties in the composition of matter after hydrogen/helium
burning prohibit a calculation of the subsequent chemical evolution of
the ocean for hydrogen-rich accretion.  However, even though the
composition is not well known, we can still use the thermal profiles as
estimates of the crust temperatures.  This is important to the evolution
of the magnetic field, as the accretion-induced heating of the crust
reduces its conductivity and hastens the Ohmic diffusion of crust
magnetic fields (\cite{gep94}).  This heating also increases the mass of
the ocean.  These effects have been considered (\cite{rom90};
\cite{gep94}; \cite{pet95}; \cite{urp95}; \cite{urp96}; \cite{kon97})
for stars accreting at $\dot{M}\lesssim 10^{-9}\msun\yr^{-1}$.

There are, however, a few neutron stars accreting globally at or near
the Eddington rate.  There are two X-ray pulsars (LMC~X-4 and SMC~X-1)
and the six bright ``Z'' sources (Sco~X-1, GX~5--1, GX~349+2, GX~17+2,
GX~340+0, Cyg~X--2).  The accreted material will have spread over the
surfaces of these star for column densities $\gtrsim 10^{14}\GramPerSc$
(see \S~\ref{s:MagneticConfinement}), so that a spherically symmetrical
approach is warranted for this calculation.  We thus use our solutions
for the thermal profile of the deep crust at accretion rates
$\mdot\sim\medd$ (see section \ref{s:core}) to estimate the Ohmic
diffusion timescales in the deep crust of these neutron stars.

\subsection{The Microphysics in the Crust}\label{s:CrustMicrophysics}

The conductivity in the crust is set by electron-phonon and
electron-impurity scattering.  In the relaxation-time approximation, the
conductivity is (\cite{yak80})
\begin{equation}
   \sigma = \frac{e^2 n_e}{m_e\sqrt{1+x^2}}\frac{1}{\nu},
\end{equation}
where $\nu$ is the sum of the electron-phonon ($\nu_{\rm ph}$) and
electron-impurity ($\nu_{\rm imp}$) collision frequencies, for which we
use the expressions from Urpin \& Yakovlev (1980),
\begin{mathletters}
\begin{eqnarray}
   \frac{1}{\nu_{\rm ph}} &\approx& \frac{\hbar^2c}{13e^2\kB T}
      \left[1+\left(\frac{\Theta}{3.5 T}\right)^2\right]^{1/2},\\
   \frac{1}{\nu_{\rm imp}} &\approx& \frac{3\pi\hbar^3}{8m_ee^4} 
      \frac{Z}{Q} \frac{1}{x}.
\end{eqnarray}
\end{mathletters}
Here $Q\equiv n_I{}^{-1}\sum_j (Z_j-\bar{Z})^2 n_j$ parameterizes the
impurities, described by charge $Z_j$ and number density $n_j$.  We
assume that $x\gg1$ and neglect anisotropies in the relaxation time due
to the magnetic field.  From these conductivities we then calculate
the local Ohmic diffusion time over a scale height,
\begin{equation}
   \label{e:DecayTime}
      \tau_{\rm diff} = 4\pi\sigma\frac{H^2}{c^2}.
\end{equation}
We are using the pressure scale height $H$ (eq.\ [\ref{e:ScaleHeight}])
as the characteristic lengthscale.  At neutron drip, $H/R\approx
0.01(2Z/A)^{4/3} (\rho/10^{11}\GramPerCc)^{1/3}$, so that 
a plane-parallel approach is valid throughout the crust.

\subsection{Ohmic Diffusion Times in the Crust}\label{s:DecayTimes}

\begin{figure}[hbp]
\centering{\epsfig{file=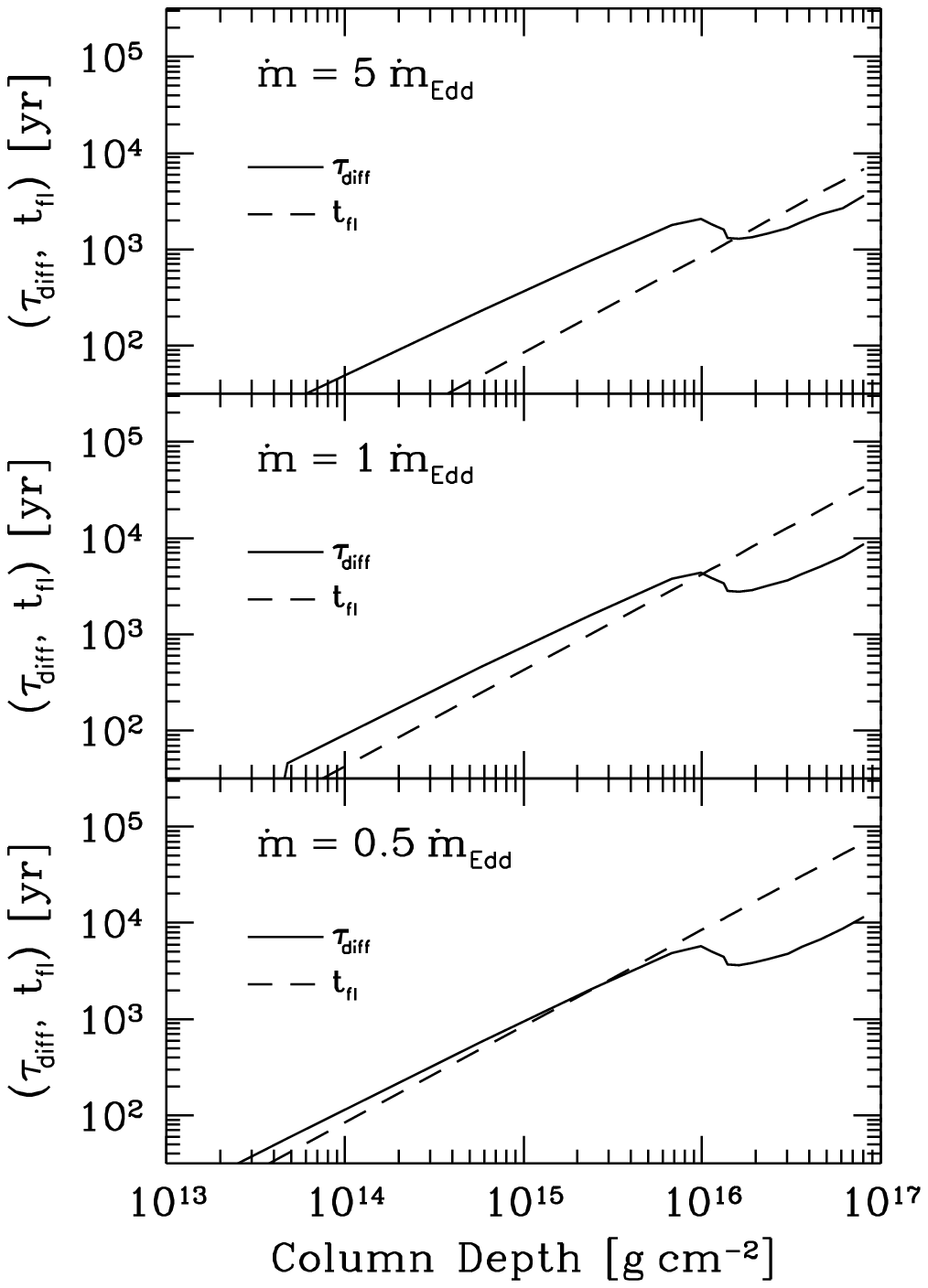,width=\hsize}}
\large\renewcommand{\baselinestretch}{0.7}\footnotesize
\caption{\protect{\footnotesize
Ohmic diffusion in the crust as a function of column depth for accretion
rates $\mdot/\medd=0.5$, 1.0, and 5.0.  The conductivity includes both
electron-phonon scattering and electron-impurity ($Q=1.0$) scattering.
We show the timescale for Ohmic diffusion ({\em solid line\/}) over a
scale height and the timescale for the crust to be pushed through a
scale height, $y/\mdot$ ({\em dashed line\/}).  The two timescales
become comparable, above neutron drip, when $\dot{M}\approx
3\ee{-9}\msun\yr^{-1}$.  Below neutron drip, the flow timescale is
always longer than the diffusion timescale for sub-Eddington accretion
rates.  The density as a function of column depth is approximately given by
eq.\ (\protect\ref{e:rho(y)}).
\label{f:OhmicDecay_Q1}}}
\end{figure}

For the temperatures in the crust, we used the estimated profiles from
\S~\ref{s:core} for the case of nonequilibrium nuclear reactions
occurring deep in the crust.  As in that section, we assume that
temperature is only a function of depth $y$, as at these depths the
accreted matter will have spread around the star.  We plot in Figure
\ref{f:OhmicDecay_Q1} ($Q=1.0$) the local Ohmic diffusion time ({\em
solid lines\/}) for accretion rates of 0.5, 1.0, and 5.0 times Eddington.
We also show the flow time over a scale height, $\tfl\equiv y/\mdot$
({\em dashed lines\/}).  A few conclusions are immediate.  First, where
the ions vibrate classically ($T\gtrsim\Theta$), the ratio $\tau_{\rm
diff}/\tfl$ is nearly independent of depth until near neutron drip;
moreover, for accretion rates $\mdot/\medd\lesssim
0.23(A/2Z)^2(T/5\ee{8}\K)$, the diffusion time is always greater than
the time for matter to flow through one scale height.  Second, impurity
scattering is unimportant throughout the crust for $Q\lesssim 1$.
Because we placed the heat sources at a fixed depth $y=y_w$ 
the thermal gradient changes sign there (see
\S~\ref{s:core}).  Electron captures remove pressure support and
therefore decrease the pressure scale height, causing the abrupt
decrease in the Ohmic timescale ({\em solid line}, Figure
\ref{f:OhmicDecay_Q1}).  Once neutron pressure dominates the equation of
state, the scale height again increases with depth.  In this region, the
flow timescale is always longer than the diffusion timescale for
$\mdot\le\medd$.

Our calculation has several differences with previous calculations.
First, the flow timescale is always less than the Ohmic diffusion
timescale, so that the downward advection of current is nonnegligible
for $\dot{M}\gtrsim 3\ee{-9}\msun\yr^{-1}$.  This agrees with the
findings of Konar \& Bhattacharya (1997), who found that the decay of
the crust field decreased as the accretion rate increased (they
considered accretion rates of less than $10^{-9}\msun\yr^{-1}$).
Second, the impurity scattering does not make a significant contribution
to the conductivity until near-nuclear densities are reached (for
$Q=1.0$).  Of course, there is no physical reason why $Q$ should have a
value near unity; in fact, it is not clear that the crust can be
adequately described as a crystalline lattice with impurities.  Given
the non-equilibrium nuclear processes in the ocean and crust, a better
description may be that of an alloy.  The chemical profile from Haensel
\& Zdunik (1990), which used $\dot{M}=10^{-10}\msun\yr^{-1}$, assumed a
single species present at each column depth, starting with a uniform
iron ocean.  This is unlikely to be realized in practice, especially if
the ocean is composed of rp-process ashes.  However, increasing the
number of species present in the crust is unlikely to {\em raise\/} the
conductivity, so that our calculation probably represents an upper bound
to the actual diffusion time.  A final difference is that our oceans are
much deeper---the melt surface is located at $y\sim5\ee{13}\GramPerSc$
($\rho\sim10^{10}\GramPerCc$).

In computing the diffusion and flow timescale, we use the local value
of the pressure scale height $H$ as the characteristic lengthscale.
In contrast, Geppert \& Urpin (1994) used a fixed lengthscale of
$832\cm$ (which is much less than the pressure scale height in the
deep crust, eq.\ [\ref{e:ScaleHeight}]).  Because $\tau_{\rm diff}$ is
proportional to the square of the lengthscale, the diffusion timescale
with their length is much shorter than ours at depths where the local
scale height exceeds their fiducial value.

The way in which these calculations are applied to the question of
global magnetic field decay depends on where one places the currents
that provide the magnetic dipole.  Geppert \& Urpin (1994) and others
have placed all of the current in the crust and argued that the surface
field will decay on the timescale given by the Ohmic diffusion time at
the depth where the majority of the current is flowing.  Given an
initial distribution of currents in the crust, the evolution of the
surface field is then determined by the competition between the decay
rate at the depth of the current and the downward advection of the
currents to higher conductivity regions (Konar \& Bhattacharya 1996).
Estimates made in this way are very sensitive to the initial current
distribution, as is clear from the change in the local Ohmic diffusion
time with depth (see Fig.\ \ref{f:OhmicDecay_Q1}).  If, however, the
field penetrates the core, then the diffusion time is roughly $R/H\sim
10\mbox{--}100$ times longer than in the absence of field penetration
(\cite{pet95}).

\section{Magnetic Confinement of Accreted Matter at a Polar Cap} 
\label{s:MagneticConfinement}

The most common sites for locally super-Eddington accretion rates are
the magnetic polar caps of accreting pulsars.  The local rates can be
quite large and most estimates of the polar cap area lead to
$\mdot\gg\medd$ at the photosphere for the majority of bright accreting
pulsars (\cite{lam73}; \cite{aro76}; \cite{els77}; \cite{bil97}).  We now
find the depth at which the matter begins to spread away from the
magnetic polar cap.  Regions underneath this spreading depth have a local
accretion rate closer to the average, $\dot{M}/4\pi R^2$.  We start by
showing that the flow on the polar cap occurs under nearly MHD
conditions for local rates $\mdot\gtrsim0.1\medd$.

\subsection{The Magnetic Reynolds Number in the Atmosphere and Ocean}
\label{s:ReynoldsNumber}
 
The magnetic pressure is always much larger than the matter pressure
until rather large depths are reached.  We thus begin in the upper
atmosphere, where $B^2\gg 8\pi P$, and ask whether the flow can cross
the field lines via Ohmic diffusion.  The fluid behavior is parameterized
by the magnetic Reynolds number ${\cal R}_m\equiv \tfl/\tau_{\rm diff}$, where
$\tau_{\rm diff}$ is the Ohmic diffusion time across a scale height
(defined in eq.\ [\ref{e:DecayTime}]).  In the absence of lateral
flow, the flow time scale over a pressure scale height is
$\tfl=y/\mdot$.  In the upper atmosphere, above and near the
hydrogen/helium burning , we estimate the conductivity as (see, e.g.,
\cite{spi62}),
\begin{equation}
   \sigma \approx  \frac{2(2\kB T)^{3/2}}{\pi^{3/2}m_e^{1/2}e^2 \Lambda},
\end{equation}
where $\Lambda\approx 8 $ is the Coulomb logarithm\footnote{We follow
the convention of Yakovlev \& Urpin (1980) and denote the Coulomb
logarithm with $\Lambda$ instead of $\ln\Lambda$} at typical atmospheric
temperatures.  We then find that the magnetic Reynolds number in the
upper atmosphere is
\begin{eqnarray}\label{e:R_mAtmosphere}
   \lefteqn{{\cal R}_m\approx }\\ \nonumber
      && 23 \left(\frac{\mdot}{\medd}\right)
	\left(\frac{10^5\GramPerCc}{\rho}\right)
	\left(\frac{T}{10^8\K}\right)^{5/2} \left(\frac{8}{\Lambda}\right).
\end{eqnarray}
When the hydrogen/helium burning is in steady-state, we can use the flux
equation (eq.\ [\ref{e:st_T}]) with Thomson scattering opacity to
eliminate $\rho$ in favor of $T$ and $\mdot$ and obtain
\begin{eqnarray}
   \lefteqn{\left.{\cal R}_m\right|_{\rm atmosphere}\approx }\\
\nonumber
   && 1400 \left(\frac{\mdot}{\medd}\right)^2 
      \left(\frac{5\times 10^8\K}{T}\right)^{1/2}
      \left(\frac{8}{\Lambda}\right), 
\end{eqnarray}
where we have used the fiducial temperature for the hydrogen/helium
burning region.  This clearly shows that, for $\mdot > 3\ee{-2}\medd$,
Ohmic diffusion is negligible prior to stable hydrogen/helium
burning.  For lower accretion rates, the matter will spread via Ohmic
diffusion prior to hydrogen/helium ignition.

Moving from the upper atmosphere to deeper regions, we now consider an
ocean composed of H/He burning ashes.  The conductivity here is mostly
set by electron-ion scattering, which we treated in
\S~\ref{s:OhmicDecay}.  For this calculation, we replace $\Lambda_{\rm ei}$
by unity.  We are interested in the case where the ocean is composed of
light elements, such as carbon, which might later ignite unstably.  We
then find that the magnetic Reynolds number in the deep ocean is
\begin{eqnarray} \label{e:R_mOcean}
   \lefteqn{\left.{\cal R}_m\right|_{\rm ocean} \approx }\\ && 291
      \left(\frac{Z}{6}\right)^{2/3} 
      \left(\frac{12}{A}\right)^{4/3} 
      \left(\frac{10^6\GramPerCc}{\rho}\right)^{1/3}
      \left(\frac{\mdot}{\medd}\right), \nonumber
\end{eqnarray}
and {\em is nearly independent of the density\/}.  As above, this serves
to define a critical accretion rate, $\mdot > 3\times 10^{-3} \medd$
(where ${\cal R}_m\sim 1$), above which we can presume nearly ideal MHD.  At
$\mdot=\medd$, the density at which carbon ignites is roughly
$10^8\GramPerCc$, which implies that ${\cal R}_m\approx 54$.  The magnetic flux
is therefore frozen into the fluid down to depths where carbon ignition
occurs for the accretion rates of interest to us.  Hence, if the magnetic
field is sufficiently strong, the fluid will remain on the polar cap
until reaching these depths.  We now estimate how strong the field must
be.

\subsection{A Simple Magnetostatic Mountain}\label{s:SimpleMountain}

Polar cap accretion causes transverse pressure gradients that are
balanced by curvature of the magnetic field.  We give an
order-of-magnitude estimate of the bending required to support a
mountain of fixed overpressure (Appendix \ref{s:GradShafranov} contains
a more detailed discussion of this problem.) In this sense, we are
envisioning accumulation of matter, followed by eventual reconnection or
Ohmic dissipation.  Our crude criterion for where spreading occurs is
where the field lines are bent by a large angle, as originally discussed
by Hameury et al.~(1983).

For simplicity, consider a poloidal, azimuthally symmetrical magnetic
field, $\vec{B}=(B_\varpi,0,B_z)$, at the polar cap.  The overpressure
due to the accumulated accretion flow distorts the field from $\vec{B} =
(0,0,B_0)$ to the perturbed configuration
$\vec{B}=(B_\varpi,0,B_0-\delta B_z)$.  Because the flow velocity for
$\mdot\sim\medd$ is so small
[$v\approx0.1\cm\second^{-1}(10^8\GramPerSc/y)$], we neglect terms
containing the velocity in the time-independent momentum equation,
\begin{equation}\label{e:Magnetostatic}
   -\grad P+\rho\vec{g}+\frac{1}{4\pi} (\curl\vec{B})\times\vec{B} = 0.
\end{equation}
We discuss the solution of equation (\ref{e:Magnetostatic}), under some
simplifying assumptions, in Appendix B.  Here, we reproduce a simple
estimate (\cite{ham83}) of the value of $B_\varpi/B_0$.  The
characteristic length in the $z$-direction is the scale height $H$ and
the characteristic length in the $\varpi$-direction is the polar cap
radius $R_{\rm cap}$.  The current density induced by the transverse
pressure gradients in the mountain is
\begin{equation}\label{e:CurrentDensity}
  \vec{J} = \frac{c}{4\pi}\curl\vec{B}
\end{equation}
and is toroidal.  The radial component of equation
(\ref{e:Magnetostatic}) implies that
\begin{equation}\label{e:RadialMagnetostatic}
   \frac{P}{R_{\rm cap}} \approx \frac{B_0}{4\pi}
   \left(\frac{B_\varpi}{H}+\frac{\delta B_z}{R_{\rm cap}}\right).
\end{equation}
The components $\delta B_z$ and $B_\varpi$ are related by
$\divr\vec{B}=0$, so $B_\varpi H \approx \delta B_z R_{\rm cap}$ and
equation (\ref{e:RadialMagnetostatic}) becomes
\begin{equation} \label{e:FieldBending}
   \frac{B_\varpi}{B_0} \approx \frac{4\pi H P}{R_{\rm cap}B_0^2}
   \left[1+ \left(\frac{H}{R_{\rm cap}}\right)^2 \right]^{-1} \approx
   \beta\frac{H}{2R_{\rm cap}},
\end{equation}
where $\beta = 8\pi P/B^2$ is the ratio of the gas and magnetic
pressures.  If $H\approx 10^3 \cm $ and $R_{\rm cap}\approx 10^5 \cm$,
then equation (\ref{e:FieldBending}) implies that the field lines become
appreciably distorted where $B_\varpi/B_0\sim 1$, or
\begin{equation}\label{e:SpreadingCriterion}
   \beta \approx \frac{R_{\rm cap}}{H}\approx 100. 
\end{equation}
Because the magnetic field lines are bent where $\beta\approx100$, we
presume that spreading occurs via reconnection or interchange
instabilities, and that lateral spreading of the matter occurs at a
pressure $P\gtrsim (B^2/8\pi)(R_{\rm cap}/H)$.  Using the relativistic
degenerate equation of state for the pressure
(eq.~[\ref{e:EquationOfState}]) and scale height (eq.\
[\ref{e:ScaleHeight}]) in equation (\ref{e:FieldBending}), we find that
the field lines will be appreciably distorted at a column
\begin{eqnarray} \label{e:spread} 
   \lefteqn{y_{\rm spread}\approx 3.4\ee{10}}\\ \nonumber
   &&  \times \left(\frac{A}{2Z}\right)^{4/5}
      \left(\frac{B_0}{10^{12}\gauss}\right)^{8/5}
      \left(\frac{R_{\rm cap}}{10^5\cm}\right)^{4/5}\GramPerSc.
\end{eqnarray}
As we show in the next section, stars accreting pure helium at
$\mdot\gtrsim 10\medd$ ignite carbon unstably before the material can
spread.

The induced current density needed to distort the field creates at the
polar cap a dipole moment,
\begin{equation}
   \vec{\mu} =
   \frac{1}{2c}\int\vec{r}'\vcross\vec{J}(\vec{r}')\,d^3r'. 
\end{equation}
From equations (\ref{e:CurrentDensity}) and
(\ref{e:SpreadingCriterion}), we see that this dipole moment opposes the
external field and scales as
\begin{equation}\label{e:InducedDipole}
   |\vec{\mu}|\sim\left(\beta\frac{H}{2R_{\rm cap}}\right) \frac{B_0R_{\rm
   cap}{}^3}{8\pi}.
\end{equation}
Though the polar cap is radially very thin, the induced dipole distorts
the magnetic field topology for a distance $R_{\rm cap}$ above the
surface and therefore can easily modify the flow of matter there.  A
totally self-consistent calculation that includes the effects of this
induced moment on the accretion flow is beyond the scope of this paper,
but it is interesting to note that mountains at these large
overpressures (if stable) can easily change the local magnetic field
strength in the polar cap region.  Whether this can cause observable
temporal changes to the pulse profile or cyclotron line energies is not
presently known.  In Appendix \ref{s:GradShafranov}, we describe our
numerical solution to equation (\ref{e:Magnetostatic}) that confirms the
estimates of the spreading criterion (eq.\ [\ref{e:SpreadingCriterion}])
and the induced dipole moment (eq.\ [\ref{e:InducedDipole}]).

\section{Unstable Thermonuclear Ignition of Carbon}
\label{s:ignition}

Carbon is the predominant element left from the steady-state (or even
unstable) burning of pure helium.  Taam \& Picklum (1978) first showed
that pure carbon oceans never burn in steady-state but rather undergo
unstable thermonuclear flashes when pycnonuclear ignition occurs at
densities $\approx (6\mbox{--}9)\times 10^9 \GramPerCc$.  For local
accretion rates near Eddington, this would result in a recurrent
explosion every ten~yr or so, as originally envisaged by Woosley \&
Taam (1976).  We now emphasize that, for high accretion rates, unstable
carbon ignition can occur at much lower densities, leading to more
frequent deflagrations.  We first discuss the carbon burning thermal
instability and then apply it to a few accreting objects.

\subsection{Thermally Unstable Carbon Burning}\label{s:CarbonBurning}

\begin{figure}[hbt]
\centering{\epsfig{file=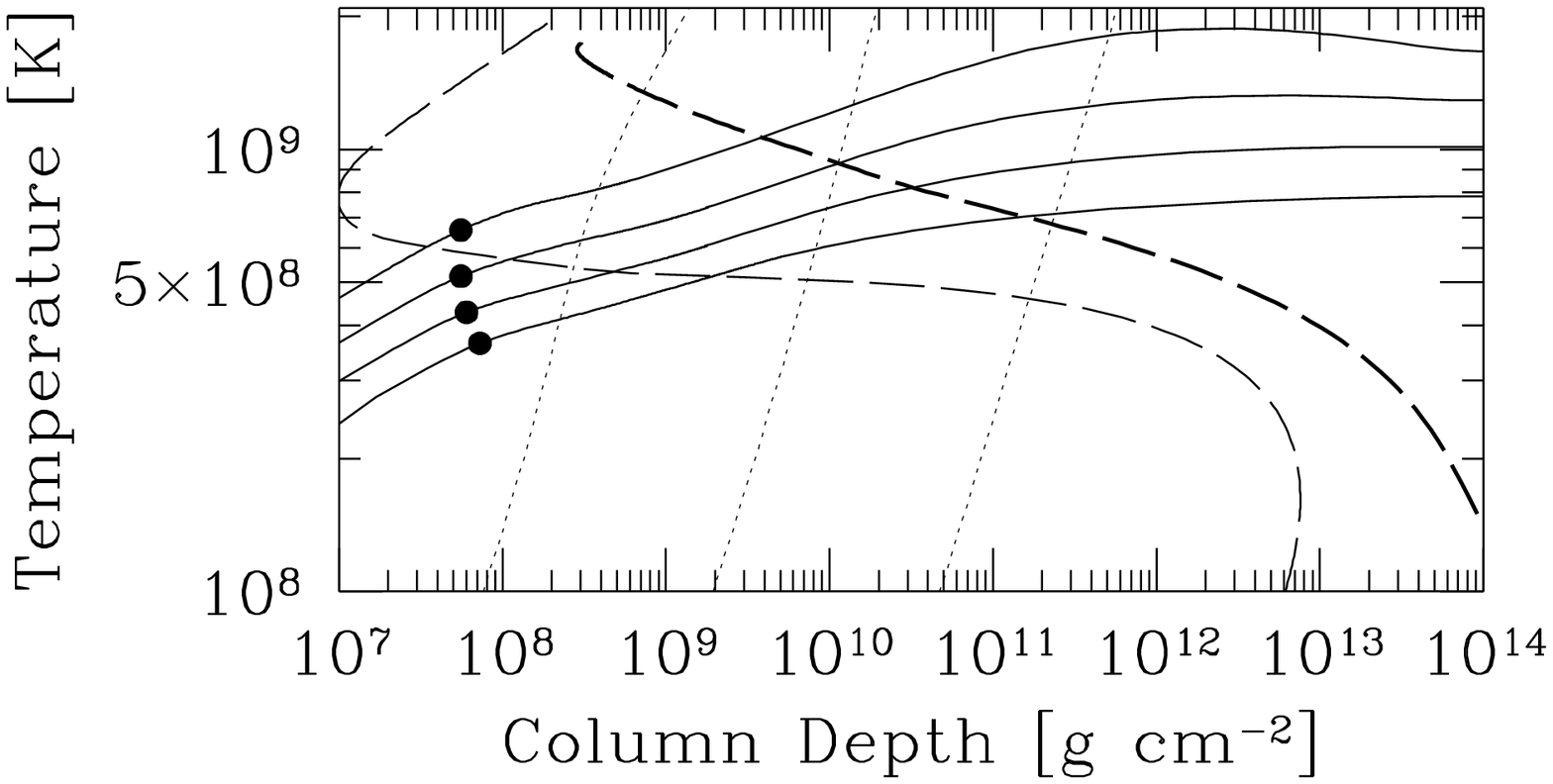,width=\hsize}}
\large\renewcommand{\baselinestretch}{0.7}\footnotesize
\caption{\protect{\footnotesize
Temperature profiles ({\em solid curves\/}) for the case of pure helium
accretion.  We take the ashes of helium burning to be pure carbon.  In
order of increasing temperature, the solutions shown are for
$\mdot/\medd = 5.0$, 10.0, 20.0, and 40.0 (see Figs.\
{\protect\ref{f:helburn10}, \protect\ref{f:helburn20},
\protect\ref{f:helburn40}, and \ref{f:HeSettle}}).  Carbon ignites where
$d\enuc/dT = d\ecool/dT$ ({\em heavy dashed curve\/}).  The filled
circles indicate where nuclear burning has decreased the abundance of
helium to half of its initial value.  To the right of the curve $\enu =
\enuc$ ({\em light dashed line\/}), carbon burning heats the ocean
faster than neutrino emission can cool it.  We show contours of constant
thermal diffusion time ({\em light dotted curves\/}) for
$\log(\tth/1\second) = 2.0$, 3.0, and 4.0.  At column depths greater
than $\sim10^{9}\GramPerSc$, the density is approximately given by eq.\
(\protect\ref{e:rho(y)}).
\label{f:ignition}}}
\end{figure}

For the \nuc{^{12}C+{}^{12}C} reaction, we use the thermonuclear
reaction rate, $\langle\sigma v\rangle$, from Caughlan \& Fowler (1988)
with a strong screening factor (\cite{oga93}).  The depth where the
energy loss from neutrino cooling equals the energy generation from
carbon burning is shown in Figure \ref{f:ignition} ({\em light dashed
curve\/}); to the right of this curve heating from carbon burning is
faster than neutrino cooling.  Carbon burning is still thermally stable
above this curve, however, since thermal conduction is so efficient.
Hence, unlike in collapsing white dwarfs and massive stellar cores, the
thermal stability of carbon ignition depends on the competition between
nuclear heating and thermal diffusion.  We therefore define the unstable
carbon ignition curve (Fig.\ \ref{f:ignition}, {\em heavy dashed
line\/}) by the equation $(\partial\enuc/\partial T)_P =
(\partial\ecool/\partial T)_P$ (\cite{fuj81}; \cite{fus87b}), where
$\ecool=\rho K T/y^2$ is a local representation of the cooling rate from
diffusion.

Upon reaching the unstable ignition curve, a fluid element containing
carbon will heat up on a nuclear burning timescale.  A preliminary
investigation of the thermal stability (both in the linear regime and a
brief series of non-linear calculations under the approximations
outlined in Bildsten 1995) determines that the thermal instability
ensues soon after carbon burning initiates.  The cooling rate from
thermal diffusion [contours of constant thermal time (Fig.\
\ref{f:ignition}, {\em light dotted lines\/}) are shown for
$\log(\tth/1\second) = 2.0$, 3.0, and 4.0] is faster than the heating
rate prior to ignition.  Once ignition occurs, the entire thermal
profile moves upwards until the heating rate greatly exceeds the cooling
rate.  Timmes \& Woosley (1992) showed that pure carbon burning fronts
move at speeds greatly in excess of $10\km\second^{-1}$ and have
intrinsic widths much less than $1\cm$.  As Bildsten (1995) noted,
combustion fronts with widths much less than a scale height propagate
very easily and are difficult to quench, as all the nuclear energy
released goes into lateral (not vertical) heating.  Thus, a local
instability should lead to ignition of all connected and combustible
carbon layers in the star and a large energy release.

\subsection{Recurrence Times and Energetics}
\label{s:Energetics}

As shown above, a pure carbon ocean under compression at high local
accretion rates initiates thermonuclear burning prior to electron
capture reactions, which occur at $\rho=3.9\times 10^{10} \GramPerCc$.
We use the asymptotically isothermal solutions developed in
\S~\ref{s:solutions} to find the ignition conditions as a function of
the local accretion rate.  This treatment is adequate, as the thermal
structure of the upper ocean is mostly determined by the steady-state
helium burning and compression, not by processes in the deep crust and
core.  In order of increasing temperature, the solutions (Fig.\
\ref{f:ignition}, {\em light solid lines\/}) are for $\mdot/\medd =
5.0$, 10.0, 20.0, and 40.0 (see Figs.\ \ref{f:helburn10},
\ref{f:helburn20}, \ref{f:helburn40}, and \ref{f:HeSettle}).  The filled
circles indicate where nuclear burning has decreased the abundance of
helium to half of its initial value.  It is the intersections of these
curves with the carbon ignition curve ({\em heavy dashed line\/}) that
sets the characteristic recurrence time, and hence the energy released
during the instability.

For example, the solution for $\mdot/\medd=5.0$ intersects the ignition
curve at $y=1.85\ee{11}\GramPerSc$ (see Fig.\ \ref{f:ignition}).  The
recurrence time is therefore $5\ee{5}\second$, and the burst energy
(assuming that fuel accumulated over $4\pi R^2$ of the surface burns) is
$1.7\ee{41}(\dot{M}/10^{-8}\msun\yr^{-1})\erg$.  The thermal time, which
sets the duration of the resulting burst in the absence of convection,
at this depth and temperature ($7.1\ee{8}\K$) is
$\tth=8.4\ee{3}\second$.  In contrast, the solution for
$\mdot/\medd=40.0$ intersects the ignition curve at
$(y=3.8\ee{9}\GramPerSc,\ T=1.1\ee{9}\K)$, so the recurrence time is
$1.3\ee{3}\second$, implying an energy release of
$4.6\ee{38}(\dot{M}/10^{-8}\msun\yr^{-1})\erg$; the thermal time here is
$\tth=470\second$.  The energy released in the instability is roughly
the energy released per gram of reactants times the amount of material
accumulated between bursts.  It is possible that not all of the
accumulated matter will be burned at once; for example, if the accretion
geometry is two disjoint polar caps, then each cap will burst
independently.  What is accurately known is the time-averaged ratio of
the energy released in quiescence to the energy released during a burst
(the $\alpha$-parameter).  For carbon burning, this ratio is
$\alpha\approx105$.

\subsection{Comparison to Observations}\label{s:Observations}

On what neutron stars can this instability occur?  In low magnetic field
objects, the matter will have spread around the star prior to carbon
ignition, in which case all of the weakly magnetic stars accreting pure
helium (4U~1820--30, 4U~1916--05, and X1850--087) will have
$\mdot<\medd$ at the depth of carbon ignition.  These stars will then
undergo ignition at such high column densities ($>10^{13}\GramPerSc$)
that the time to reach ignition will be $\sim10\mbox{--}100\yr$,
resulting in released energies of order
$10^{44}\erg(\dot{M}/10^{-8}\msun\yr^{-1})(t_{\rm rec}/10\yr)$.  No such
events have ever been identified.  This is not surprising, given the
long recurrence time and short duration of the event.

Another possible binary with helium-rich accretion at high accretion
rates is Cygnus~X-3.  van Kerkwijk et al.~(1996) have recently tightened
the case that this X-ray binary consists of a WN Wolf-Rayet star
orbiting a compact object (\cite{van92}; \cite{van93}).  Neither
pulsations nor type I X-ray bursts have ever been seen from this object.
Detailed X-ray timing and spectral studies with {\em EXOSAT\/}
(\cite{ber95}; \cite{smi96}) did not allow for a simple classification
of this object in terms of the standard accreting neutron star
phenomenology (i.e., Z and atoll sources).  We would expect the helium
burning to be unstable if the neutron star is accreting nearly pure
helium at a rate less than ten times the Eddington limit.  It might be
possible that the type I bursts are obscured by the well-documented
corona about this object or that the compact object is a black hole.
The carbon burning would be unstable on this object and potentially have
shorter recurrence times of order years.

The most intriguing scenario for our work is an accreting X-ray pulsar,
where the high magnetic field confines the accreted matter onto a small
fraction of the stellar surface.  The enhancement of the local accretion
rate then decreases the recurrence time, possibly to values as short as
a day, even at globally low accretion rates.  The methodology by which
unstable nuclear burning on an accreting X-ray pulsar is used to place
constraints on the surface magnetic field and polar cap size is
discussed for GRO~J1744--28 in Bildsten \& Brown (1997).

For one pulsar, 4U~1626--67, our calculations are directly applicable.
Chakrabarty et al.~(1997) inferred a global accretion rate in excess of
$2\ee{-10}\,\msun\yr^{-1}$, and the constraints on the orbital
parameters ($P_{\rm orbit}=42\min$) imply that the accreting companion
is a helium-rich star (\cite{cha98}).  This pulsar rotates at $P_{\rm
spin}=7.68\second$ and has a corotation radius of $6.5\ee{8}\cm$.  If
we presume that the magnetospheric radius is near equilibrium, so that
the accretion follows the field lines from the corotation radius
(\cite{lam73}; \cite{gho79}), then the local accretion rate at the polar
cap (each polar cap has area $10^{10}\cm^2$) is $10^6 \localrate$.
Carbon ignition then occurs at a column depth $y_{\rm ign}\approx
1.9\ee{10}\GramPerSc$, which the accreted material reaches within $4\hour$
of deposition onto the surface.  The ignition occurs long before
the matter has spread around the star for $B\gtrsim8\ee{11}\gauss$.  The
thermal time at that depth is roughly $100\second$.  These quantities set
the recurrence and duration of unstable carbon burning for this
particular polar cap model.  Unfortunately, the recurrence time is very
sensitive to the local accretion rate (and hence is a sensitive function
of the accretion geometry and global rate).  In this case it is better
to invert the problem: given a recurrence time, we can immediately infer
the local accretion rate, and thus the polar cap area.  From Figure
\ref{f:ignition}, the recurrence time is roughly
$6\ee{4}(10.0\medd/\mdot)^{2.84}\second$.  There have been many reports
of flares on shorter timescales from this pulsar (recurrence times of
1000 seconds).  These flares could result from carbon burning if the
matter is either confined to an area smaller (roughly $2\ee{9}\cm^2$)
than our naive estimate, or if the global accretion rate is a factor of
ten higher than the minimum stated earlier, as suggested by the recent
distance estimate of Chakrabarty (1998).  If the polar cap is larger
than $10^{10}\cm^2$, as suggested by accretion models in which the
matter penetrates the magnetopause via a Rayleigh-Taylor instability
(\cite{aro76}; \cite{els77}), then these flares could be caused by unstable
helium burning.  The recurrence time would then imply a local accretion
rate of order $10^5\localrate$ (low enough for helium burning to be
unstable), or a polar cap area $A_{\rm cap}\sim 10^{11}\cm^2$, which
is still much less than the total surface area.

We were initially very interested in the large daily flares 
from the X-ray pulsar LMC X-4 (\cite{lev91}; Dal Fiume et al.~1997).  During
these $\sim 30{\rm\,min}$ flares, the pulsed fraction increases, the
pulse becomes single peaked and sharply defined, and the spectrum
softens.  The ratio of the persistent fluence to the time-averaged burst
fluence (the $\alpha$ parameter) is $\sim 33$; hence, the energetics of
the flares are reasonably consistent with the unstable burning of
accumulated matter.  The time-averaged luminosity of $\approx
4\ee{38}\erg\second^{-1}$ implies a super-Eddington {\em global\/}
accretion rate of approximately $4\ee{-8}\msun\yr^{-1}$, and so we
investigated the possibility that the flares could be from carbon
burning.  The difficulty is that the massive companion is providing
hydrogen-rich material, in which case we do not expect large amounts of
carbon in the deep ocean, as the hydrogen burning is known to produce
much heavier ashes.  It presently seems unlikely that carbon
burning is the cause of these flares.

Even though we cannot presently explain them, we would like to note
three other accreting pulsars that exhibit large flares: SMC~X-1,
GX~304--1, and EXO~2030+375.  {\em EXOSAT\/} observations of SMC~X-1
(Angelini, Stella, \& White 1991) show an $\sim 80\second$ flare with a
rise time $\sim 1\second$.  This behavior is qualitatively duplicated in
GX~304--1 (\cite{mcc77}), which has flares of duration $\sim 100\second$
and rise time $\sim 10\second$.  The flares of EXO~2030+375
(\cite{par89}) are characterized by both a longer rise time ($\sim
30\minute$) and a longer duration ($\sim 1\mbox{--}2\hour$).
Moreover, the recurrence time is $\sim 4\hour$.  Because EXO~2030+375 is
a transient X-ray pulsar, it is expected that the accretion rate varies
greatly in time.  With the launch of {\em RXTE\/}, it is possible to
obtain detailed pulse and burst profiles from these sources.  For
example, the sharpening of the pulse profile on LMC~X-4 during a burst
would be expected from a single polar-cap burst.  If this is true, then
it should be possible to determine if only one polar cap bursts, or if
the polar caps alternate.  These types of observations would help in the
eventual unraveling of the cause of these flares.

\section{Summary and Conclusions}\label{s:conclusions}

 We have investigated the thermal and compositional structure of a
rapidly accreting neutron stars in many different environments, both
highly and weakly magnetic.  We first demonstrated, in
\S~\ref{s:SteadyStateHeBurning}, that for accretion rates
$\mdot\gtrsim10.0\medd$ (for which pure helium burning is stable), the
ashes of pure helium burning are mostly carbon (Figs.\
\ref{f:helburn10}, \ref{f:helburn20}, and \ref{f:helburn40}).  In
\S~\ref{s:CarbonBurning}, we calculated the unstable carbon
ignition curve; the intersection of our settling solutions with that
curve (see Fig.\ \ref{f:ignition}) determine the recurrence times and
energetics of the unstable carbon burning.  For the high local
accretion rates underneath a pulsar polar cap, we found that episodes
of unstable burning can recur within hours to days.  This is much
shorter than the previously found recurrence times of years for lower
accretion rates.  In \S~\ref{s:Observations}, we listed some promising
sources for observing such unstable carbon burning.  We are especially
intrigued by the $1000\second$ flares on 4U~1626--67, which accretes
from a helium-rich star.  There have also been large flares seen from
LMC~X-4, SMC~X-1, GX~304--1, and EXO~2030+375.  Further observations
of the burst energetics, the recurrence time, the burst duration, and
the changes in the spectrum, the pulse profile and phase, and the
pulsed fraction of the luminosity during a burst should reveal the
nature of these events.

Accreting X-ray pulsars offer a promising laboratory for studying the
effects of a locally high accretion rate.  In the upper atmosphere and
ocean, the Ohmic diffusion time is always much longer than the flow
timescale (\S~\ref{s:ReynoldsNumber}), so that the matter is tied to the
field lines.  We showed, in \S~\ref{s:SimpleMountain}, that the magnetic
field keeps matter from spreading (which occurs when the initially
vertical magnetic field is bent so that the poloidal component is
$B_\varpi\sim B_0$) around the star until the gas pressure is $P\approx
(R_{\rm cap}/H)B^2/8\pi$.  Where the gas pressure is sufficient to bend
the field lines, a dipole moment of order $B_0R_{\rm cap}{}^3$ is
induced (see Appendix \ref{s:GradShafranov}).  This induced dipole is in
opposition to the stellar dipole and can affect the magnetic field
strength, and consequently the accretion stream, at the polar cap.  If
time dependent, this could lead to observable changes in the pulse
profiles or cyclotron line energies from these objects.

We also discussed the case of hydrogen and helium burning
(\S~\ref{s:SteadyStateHHeBurning}) for accretion rates where the burning
is stable.  We described in detail the thermal state of the ocean,
crust, and core for neutron stars accreting globally at rates in excess
of $\sim 10^{-9} \msun\yr^{-1}$.  In \S~\ref{s:OhmicDecay}, we used
these thermal profiles to estimate the Ohmic diffusion time throughout
the star.  We found that the hotter crust of a rapidly accreting neutron
star shortens the diffusion time to less than $10^8\yr$ throughout the
crust for an Eddington accretion rate.  Moreover, for accretion rates
$\mdot\sim\medd$, the flow time is less than the diffusion time
throughout the crust.  This raises intriguing questions about the
evolution of global magnetic fields in rapidly accreting neutron stars,
like the bright ``Z'' sources.

\acknowledgements

It is a pleasure to thank Jon Arons and Dana Longcope for numerous
discussions about magnetohydrostatics, Ron Taam and Michael Wiescher for
informing us about deep hydrogen burning, A. Khokhlov for notes on the
electron scattering opacity, Frank Timmes for information on the
\nuc{^{12}C\,(\alpha,\gamma)\,^{16}O} reaction, and D. Bhattacharya,
L. Hernquist, and A. Melatos for comments on the manuscript.  This work
was supported by NASA via grants NAG 5-2819 and NAGW-4517 and by the
California Space Institute (CS-24-95).  E. F. B. received support from a
NASA GSRP Graduate Fellowship under grant NGT-51662, and
L. B. acknowledges support as an Alfred P. Sloan Foundation fellow.

\appendix
\section{Transformation to Column Depth Coordinates}\label{s:isobar}

There is a subtlety in the plane-parallel equations for subsonic
accretion flow: namely, whether constant pressure contours (isobars) are
spatially fixed or move radially outward during accretion.  (For a
brief discussion, see \S~3.2 of Bildsten, Salpeter, \& Wasserman 1992.)
For example, accretion of an incompressible liquid would lead to outward
motion of the isobars (relative to the stellar center of mass) at a rate
$v_{\rm iso} = \mdot/\rho$.  In this appendix we show that allowing for
compression implies that the isobars have constant altitudes on
timescales shorter than that needed for the accreted matter to reach
pressures where the equation of state becomes nearly incompressible.  For
neutron stars this timescale is greater than $10^6\yr$, i.e., the
equation of state is very compressible until the internuclear spacing
is on the order of a nuclear diameter.

The transformation equations between plane-parallel and column depth
coordinates are
\begin{mathletters}
\begin{eqnarray}
\left.\frac{\partial}{\partial z}\right|_t 
   &=& -\rho\left. \frac{\partial}{\partial y} \right|_t\\
\left.\ddt{}\right|_y 
   &=& \left.\ddt{}\right|_z + \left.\ddt{z}\right|_y
   \frac{\partial}{\partial z} \nonumber\\
   &\equiv& \left.\ddt{}\right|_z + v_{\rm iso} \frac{\partial}{\partial z}.
\end{eqnarray}
\end{mathletters}
The term $(\partial z/\partial t)_y$ is the upward rate of motion of an
isobar $v_{\rm iso}$.  For example, consider a polytropic gas of index
$n$, $P\propto \rho^{1+1/n}$, sitting on an incompressible floor at
$z=0$, and let quantities evaluated at $z=0$ be denoted by a subscript
$0$.  Hydrostatic balance, $dP/dz = -\rho g$, implies (we are
neglecting the ram pressure)
\begin{equation}
   \rho = \rho_0\left(1 - \frac{z}{\lambda}\right)^n,
\end{equation}
where $\lambda\equiv g\rho_0/[(n+1) P_0]$ is the total height of
the polytropic atmosphere.  The column depth $y=y(z,t)$ is
\begin{equation} \label{e:polycol}
   y = \lambda\rho_0\frac{1}{n+1}
   \left(1-\frac{z}{\lambda}\right)^{n+1} =
   \frac{P_0}{g}\left(1-\frac{z}{\lambda}\right)^{n+1}.
\end{equation}
Now, $P_0/g$ equals the accumulated column mass of the atmosphere,
$y_0$.  Accretion will clearly increase $y_0$ ($dy_0/dt =
\mdot$).  Differentiating equation (\ref{e:polycol}) with respect to $t$
while holding $y$ fixed yields
\begin{equation}
   v_{\rm iso} = \frac{\mdot}{\rho_0}, 
\end{equation}
so that the velocity of a fluid element is 
\begin{equation}
   v = -\frac{\mdot}{\rho} + v_{\rm iso}
   = -\frac{\mdot}{\rho}\left(1-\frac{\rho}{\rho_0}\right).
\end{equation}
The advective derivative is then simply 
\begin{eqnarray}
   \ddt{} + v\ddz{} = 
   \ddt{}+ \mdot\left(1-\frac{\rho}{\rho_0}\right)\ddy{}\nonumber\\
   \approx \ddt{}+\mdot\ddy{}
\end{eqnarray}
far above the incompressible region where $\rho \ll \rho_0$. 

\section{Magnetic Confinement of the Accreted Matter: A More Detailed
Approach}\label{s:GradShafranov}

To investigate the magnetic confinement of a mound of accreted matter in
more detail, we considered the following idealized problem.  In a
cylindrical geometry $(\varpi,\phi,z)$, we parameterize the magnetic
field by a function $\psi(\varpi,z)$ such that
\begin{equation} \label{e:Bpar}
   \vec{B} = \frac{1}{\varpi}\grad\psi\times \vec{e}_\phi,
\end{equation}
where $\vec{e}_\phi$ is the basis vector in the azimuthal direction and
$\psi(0,z)=0$.  We are assuming that the magnetic field is poloidal,
i.e., $\vec{B}=(B_\varpi,0,B_z)$.  This parameterization of $\vec{B}$
ensures that $\divr\vec{B}=0$ is satisfied identically and that the flux
$\Phi$ crossing a disk of radius $R$ in a plane of constant $z$ is
$\Phi=2\pi\psi(R,z_0)$.  As a result, field lines are tangent to lines
of constant $\psi$.  This technique is well-known in solar physics,
where this approach (with some additional assumptions about the nature
of the toroidal field) leads to the {\em Grad-Shafranov\/} equation
(see, e.g., \cite{fre87}), a variant of which we now derive.

Hameury et al.~(1983) performed a similar calculation and concluded
that no significant bending of the field lines occurred.  This
approximation is only true if $\beta\left(H/R_{\rm cap}\right)^2 \ll 1$,
where $H$ is the height of the region.  In their numerical calculations,
Hameury et al.~(1983) used integration heights of 30 and 70~m; the
corresponding values of $[\beta (H/L)^2]_{z=0}$ were $\approx 0.7$ and
$\approx 60$.  Thus their approximations were not valid at the bottom of
the region of integration, where the greatest deformation of the field
lines occurs.

Using the parameterization in equation (\ref{e:Bpar}), we transform
$\curl\vec{B}$ into
\begin{equation}
   \curl\vec{B} = -\vec{e}_\phi \frac{\partial }{\partial
   	\varpi}\frac{1}{\varpi} \frac{\partial \psi}{\partial \varpi} + 
	\frac{\partial^2\psi}{\partial z^2} \equiv -\vec{e}_\phi
   	\frac{1}{\varpi}\triangle\psi.
\end{equation}
In this problem, we are considering the gas to be polytropic,
$P=\Pi\rho^{(1+1/n)}$, so that the enthalpy is $w=\int
dP/\rho=(n+1)P/\rho$.  Using this to eliminate the pressure in equation
(\ref{e:Magnetostatic}) and substituting the parameterization of \vec{B}
(eq.\ [\ref{e:Bpar}]), we have
\begin{equation}\label{e:TransformedMagnetostatic}
   \rho\grad(w+\gp)+\frac{1}{4\pi\varpi^2}\grad\psi\triangle\psi = 0,
\end{equation}
where $\gp$ is the Newtonian potential.  The component of equation
(\ref{e:TransformedMagnetostatic}) parallel to a field line is
$F(\psi)\equiv w(\psi,z)+\gp(z)=\;\mbox{constant}$.  Inverting this
equation then gives the density as a function of $\psi$ and $z$,
\begin{equation}\label{e:density}
   \rho(\psi,z) = \left\{\left[\frac{1}{(n+1)\Pi}\right]
      \left[F(\psi)-\gp(z)\right] \right\}^n.
\end{equation}
Thus, $F(\psi)$ specifies how each field line crossing the boundary of
the cylinder is loaded with matter.  The component of equation
(\ref{e:TransformedMagnetostatic}) perpendicular to a field line implies
that
\begin{equation}\label{e:GradShafranov}
   \rho\frac{dF}{d\psi} + \frac{1}{4\pi\varpi^2}
   \left(\triangle\psi\right) = 0.
\end{equation}
The system of equations (\ref{e:density}) and (\ref{e:GradShafranov}) is
then closed by specifying $F(\psi)$ and boundary conditions [at
$\varpi=(0,R_{\rm cap})$ and $z=(0,h)$] on $\psi$.  We choose $\psi$ so
that, along the boundary, the magnetic field matches a uniform field,
\begin{equation}
   \psi = \frac{1}{2}B_0\varpi^2.
\end{equation}
In other words, all of the field lines enter and leave the computational
box and are presumed to match onto a uniform field above and below the
region of integration.  The choice of $F(\psi)$ is somewhat arbitrary in
the absence of detailed information about the accretion flow.  A
convenient parameterization is
\begin{equation}
   F = F_0\left[\exp\left(-a\left|\frac{\psi}{B_0 R_{\rm
   cap}^2}-u\right|\right) + d\right],
\end{equation}
with this choice we can simulate either a hollow or filled polar cap
geometry.  In addition, we require that the solutions have a topology
where all field lines exit the boundaries of the region.  This ensures
that $F(\psi)$ reflects the field configuration for matter arriving from
$z\rightarrow\infty$.

As a test of the crude scaling argument (eq.\ [\ref{e:FieldBending}]),
we solved equation (\ref{e:GradShafranov}) for a region of one scale
height with $F=F_0[\exp(-\psi/B_0R_{\rm cap}^2)+1]$.  The top of the region was
chosen to have fixed density, $\rho=2\ee{6}\GramPerCc$.  We varied
independently the scale height (by changing the surface gravity), the
radius of the polar cap, and the magnetic field strength.  We chose
values for which the distortion of the field was slight.  The value of
$B_\varpi/B_0$ scaled linearly with $h$ and $L$, and inversely as
$B_0{}^{-2}$, in agreement with equation (\ref{e:FieldBending}).

\begin{figure}[hbt]
\centering{\epsfig{file=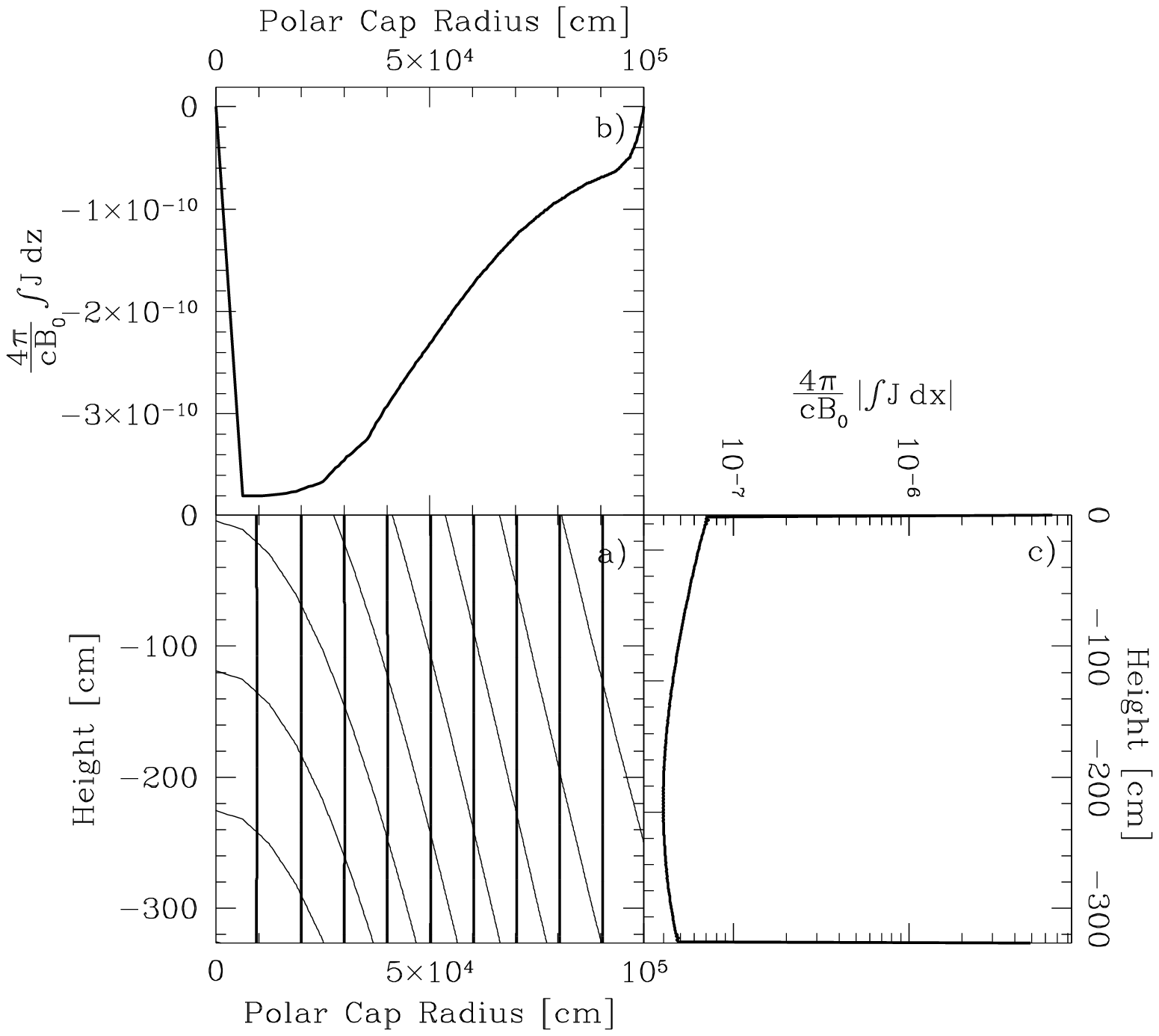,width=\hsize}}
\large\renewcommand{\baselinestretch}{0.7}\footnotesize
\caption{\protect{\footnotesize
The confinement of accreted matter by a magnetic field.  In panel {\em
a\/} we show isochores ({\em thin lines\/}) for a matter distribution,
at $z\rightarrow\infty$, of $F\propto[1 + \exp(-|\varpi/R_{\rm
cap}|^2)]$, where $\varpi$ is the coordinate along the surface.  The
magnetic field ({\em heavy lines\/}) is assumed to be uniform in the
region outside the plot. In panels {\em b\/} and {\em c\/} we show the
current density integrated over $z$ and $\varpi$, respectively.  The
parameter $\beta(H/R_{\rm cap})$ varies from 0.24 at the top to 0.37 at
the bottom, so that no appreciable spreading occurs, even though
$\beta=8\pi P/B^2=74$ at the top ($z=0$).  The induced dipole moment
(not including the current sheets at the top and bottom) is $\vec{\mu} =
-0.078 B_0R_{\rm cap}{}^3\vec{e}_z$.
\label{f:Confinement}}}
\end{figure}

\begin{figure}[hbt]
\centering{\epsfig{file=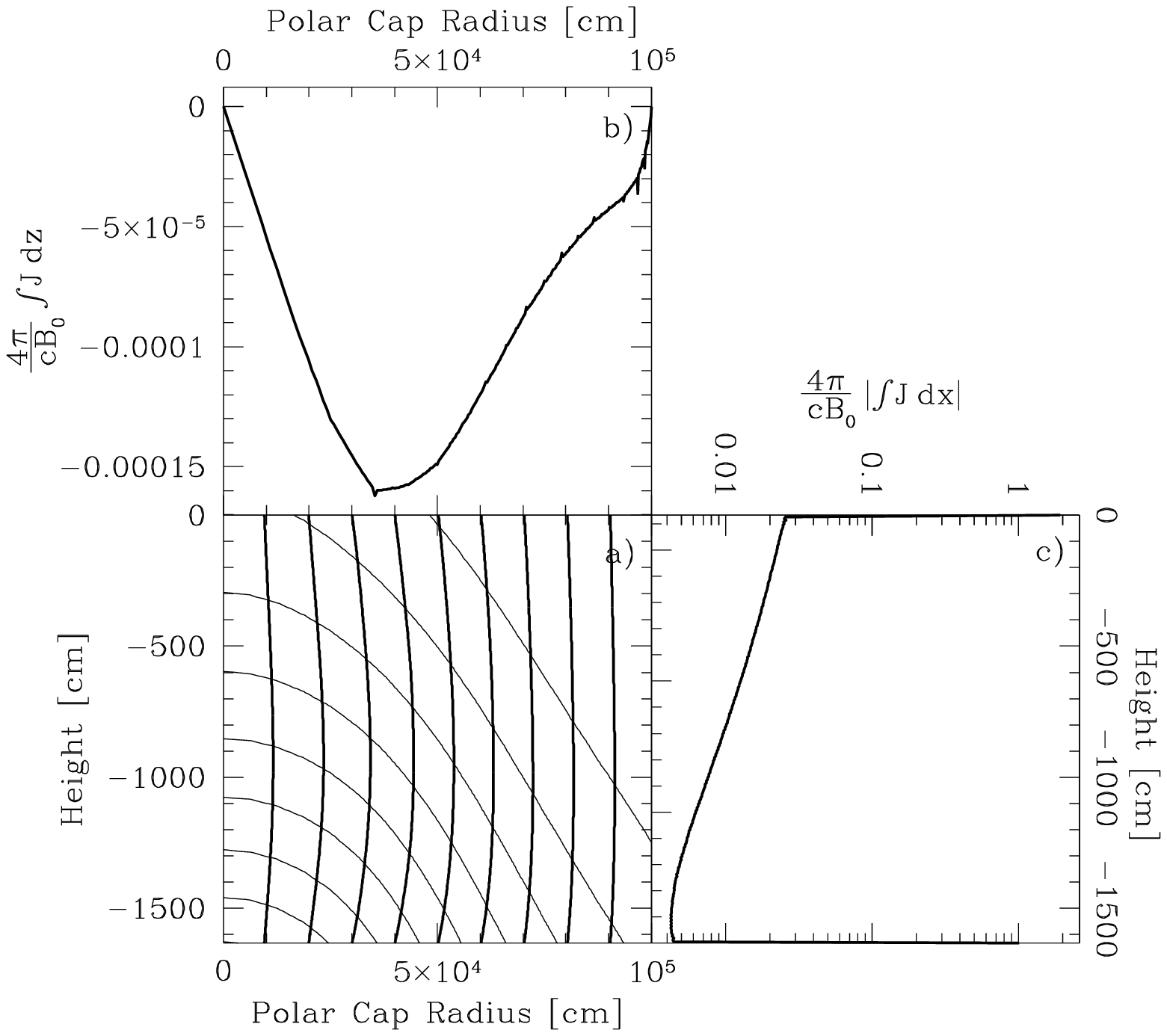,width=\hsize}}
\large\renewcommand{\baselinestretch}{0.7}\footnotesize
\caption{\protect{\footnotesize
Same as Fig.\ \protect\ref{f:Confinement}, except that the matter
distribution at $z\rightarrow\infty$ is
$F\propto[0.5+\exp(-|\varpi/R_{\rm cap}|^2)]$, and $\beta(H/R_{\rm
cap})$ varies from 0.97 to 6.1.  The displacement of the field line from
its unperturbed configuration is about $500\cm$, so that $B_\varpi/B_0$
is of order unity, as expected (note that the vertical and horizontal
scales in panel {\em a\/} are greatly different, so that the field line
distortion is not as apparent).  The induced dipole moment (not
including the current sheets at the top and bottom) is $\vec{\mu} =
-0.82 B_0R_{\rm cap}{}^3\vec{e}_z$.
\label{f:Spreading}}}
\end{figure}

In Figures \ref{f:Confinement} and \ref{f:Spreading} we show two
numerical solutions to equation (\ref{e:GradShafranov}).  Panel {\em a\/}
depicts the constant density contours ({\em thin lines\/}) and the lines
of constant flux ({\em thick lines\/}).  In panels {\em b\/} and {\em
c\/} we show 
the integral of $J$ over the vertical coordinate $z$ and the radial
coordinate $\varpi$, respectively.  The matter overpressure induces a
dipole moment, opposed to the stellar field, of order $B_0 R_{\rm
cap}{}^3$ inside the region.  The existence of large current sheets at
the top and bottom of the region is evident in panel {\em c}.  These sheets
are required by the boundary conditions that the field be uniform
outside the region.  In Figure \ref{f:Confinement} (Fig.\ \ref{f:Spreading}),
the parameter $\beta(H/R_{\rm cap})$ varies from 0.24 (0.97) at the top
to 0.37 (6.1) at the bottom.  As can be seen in Figure
\ref{f:Spreading}{\em a}, the maximum displacement of a field line from its
equilibrium location is of order $500\cm$, so that $B_\varpi/B_z$ is of
order unity.  (The field lines are bent more than is visually apparent
from panel {\em a\/} due to the different scales for the vertical and lateral
displacements).  The induced dipole (not including the current sheets at
the top and bottom) is $\vec{\mu} = -0.078 B_0 R_{\rm cap}{}^3\vec{e}_z$
($\vec{\mu} = -0.82 B_0 R_{\rm cap}{}^3\vec{e}_z$) for Figure
\ref{f:Confinement} (Fig.\ \ref{f:Spreading}).

While the field lines threading the crust are anchored in place, the
field lines through the outer atmosphere are free to move outward.
Because of the large induced dipole, the field lines in the atmosphere
will be distorted from the matter-free configuration for a distance
$\sim R_{\rm cap}\gg H$ above the ocean.

\end{document}